\documentclass[12pt]{article}

\usepackage{times}
\usepackage{amssymb}
\usepackage{amsmath}
\usepackage{array}
\usepackage[margin=2.5cm]{geometry}
\usepackage{fancyhdr}
\usepackage{cite}
\usepackage[format=hang,=small,labelfont=bf]{caption}
\usepackage{authblk}

\usepackage{color}
\usepackage{pdfcolmk} 
\usepackage{setspace}
\usepackage{soul}
\usepackage[hidelinks]{hyperref}

\ifx\pdfoutput\undefined
\usepackage{graphicx}
\else
\usepackage[pdftex]{graphicx}
\usepackage{epstopdf}
\fi

\usepackage{pdflscape} 

\usepackage[version=3]{mhchem} 

\newcolumntype{x}[1]{>{\raggedright\hspace{0pt}}p{#1}}

\graphicspath{{.},{./f/}}

\newcommand{\etal}{\textit{et al.{}}}

\newcommand{\abin}{\textit{ab initio}}
\newcommand{\CoPt}{Co--Pt}

\title{The phase stability of large-size nanoparticle alloy catalysts at \textit{ab initio} quality using a nearsighted force-training approach}

\author{Cheng Zeng$^{\dag}$, Sushree Jagriti Sahoo$^{\ddag}$, Andrew J. Medford$^{\ddag}$, Andrew A. Peterson$^{\dag, \text{*}}$}
\affil{$\text{\dag}$School of Engineering, Brown University, Providence, Rhode Island, 02912, United States.\\$\text{\ddag}$ School of Chemical and Biomolecular Engineering, Georgia Institute of Technology, Atlanta,
Georgia 30318, United States.\\$\text{*}$Corresponding author: Email: andrew\_peterson@brown.edu, Tel: +1 401-863-2153\\}

\begin{document}
\maketitle


\begin{abstract}
Co--Pt alloyed catalyst particles are integral to commercial fuel cells, and alloyed nanoparticles are important in many applications.
Such systems are prohibitive to fully characterize with electronic structure calculations due to their relatively large sizes of hundreds to thousands of atoms per simulation, the huge configurational space, and the added expense of spin-polarized calculations.
Machine-learned potentials offer a scalable solution; however, such potentials are only reliable if representative training data can be employed, which typically also requires large electronic structure calculations.
Here, we use the  nearsighted-force training approach that allows us to make high-fidelity machine-learned predictions on large nanoparticles with $>$5,000 atoms using only small and systematically generated training structures ranging from 38--168 atoms.
The resulting ensemble model shows good accuracy and transferability in describing the relative energetics for Co--Pt nanoparticles with various shapes, sizes and Co compositions.
It is found that the fcc(100) surface is more likely to form a L1$_0$ ordered structure than the fcc(111) surface.
The energy convex hull of the icosahedron shows the most stable particles have Pt-rich skins and Co-rich underlayers, and is in quantitative agreement with one constructed by brute-force first-principles calculations.
Although the truncated octahedron is the most stable shape across all sizes of Pt nanoparticles, a crossover to icosahedron exists due to a large downshift of surface energy for CoPt nanoparticle alloys.
The downshift can be attributed to strain release on the icosahedron surface due to Co alloying.
We introduced a simple empirical model to describe the role of Co alloying in the crossover for Co--Pt nanoparticles.
With Metropolis Monte-Carlo simulations we additionally searched for the most stable atomic arrangement for a truncated octahedron with equal Pt and Co compositions, and also we studied its order--disorder phase transition.
We validated the most stable configurations with a new highly scalable density functional theory code called SPARC.
From the outermost shell to the center of a large Co--Pt truncated octahedron, the atomic arrangement follows a pattern: Pt $\rightarrow$ Co $\rightarrow$ L1$_2$(Pt$_3$Co) $\rightarrow$ L1$_2$(PtCo$_3$) $\rightarrow$ L1$_0$(PtCo) $\rightarrow$ $\cdots$ $\rightarrow$ L1$_0$(PtCo).
Lastly, the order--disorder phase transition for a Co--Pt nanoparticle  exhibits a lower transition temperature and a smoother transition, compared to the bulk Co--Pt alloy.
\end{abstract}


\newpage

{\singlespace \footnotesize \noindent \textbf{Keywords:} Co--Pt nanoparticles, Nearsighted force training, Global optimization,  Morphology crossover, Order--disorder transition}

TOC Graphic:
\begin{center}\includegraphics[width=3.33in]{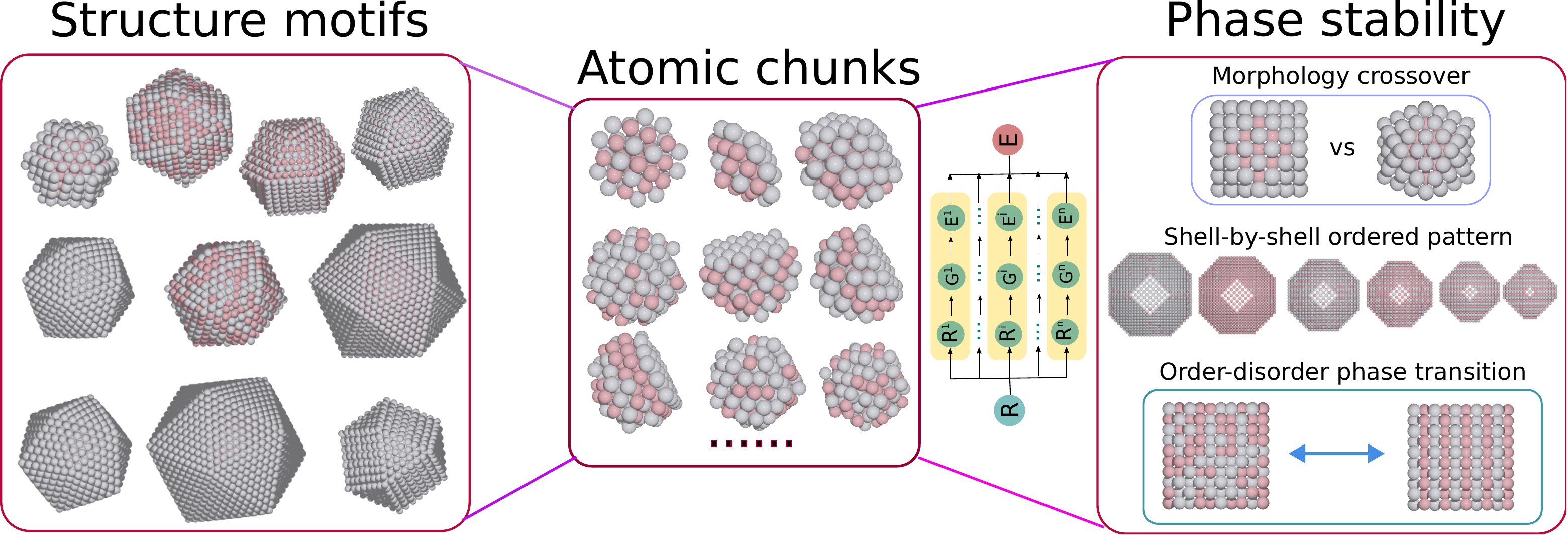}\end{center}




\newpage

\section{Introduction}

Bimetallic nanostructures have received increasing attention in the past two decades~\cite{Ristau1999, Sun2000, Ferrando2008}.
Among the family of bimetallic nanostructures, Co--Pt nanoparticles have a wide range of applications in heterogeneous catalysis and magnetic storage~\cite{Li2020, Sun2004, Jun2007, Christodoulides2000}.
Co--Pt nanoparticles have been commercialized as electrocatalysts in fuel-cell electric vehicles such as Toyoto Mirai~\cite{Yoshida2015, Wang2021}.

The size, composition, shape, and orderliness of \CoPt\ nanoparticles all play a pivotal role in controlling the structure and hence the chemical and physical properties, including the catalytic activity.
We provide several examples below, and note they are not all in agreement with one another.
For example, Li \etal\ have shown that a core-shell Co--Pt nanoparticle with an ordered core loses fewer Co atoms in electrocatalysis applications than that with a random core~\cite{Li2020}.
Others have shown the magnetic and optical properties are closely associated with the orderliness of the structure~\cite{Mehaddene2004, Andreazza2015}.
Yang \etal\ suggested, based on Monte-Carlo simulations, that disordering is initiated at the surface due to the reduced coordination, hence lowering the ordering temperature~\cite{Yang2006}.
M{\"u}ller and Albe pointed out that surface segregation of one element can have a large impact on the ordering~\cite{Muller2005}.
Alloyeau \etal\ showed in both experiments and simulations that the shape and size affect the order--disorder transition, indicating that a larger nanoparticle tends to show a higher transition temperature and the size effect is uniquely determined by the smallest length of a nanoparticle~\cite{Alloyeau2009}.
Alloyeau's large-scale simulation was based on tight-binding potentials fitted to experimental and first-principles calculations, and the most stable structure for an equal-composition nanoparticle was identified to be a fully L1$_0$ ordered truncated octahedron~\cite{Alloyeau2009}.
However, first-principles calculations by Gruner \etal{} suggested that the most stable shape of a \CoPt\ nanoparticle at small sizes is not the L1$_0$ ordered regular truncated octahedron, but a multiply twinned icosahedron.

\begin{figure}
\centering
\includegraphics[width=5.in]{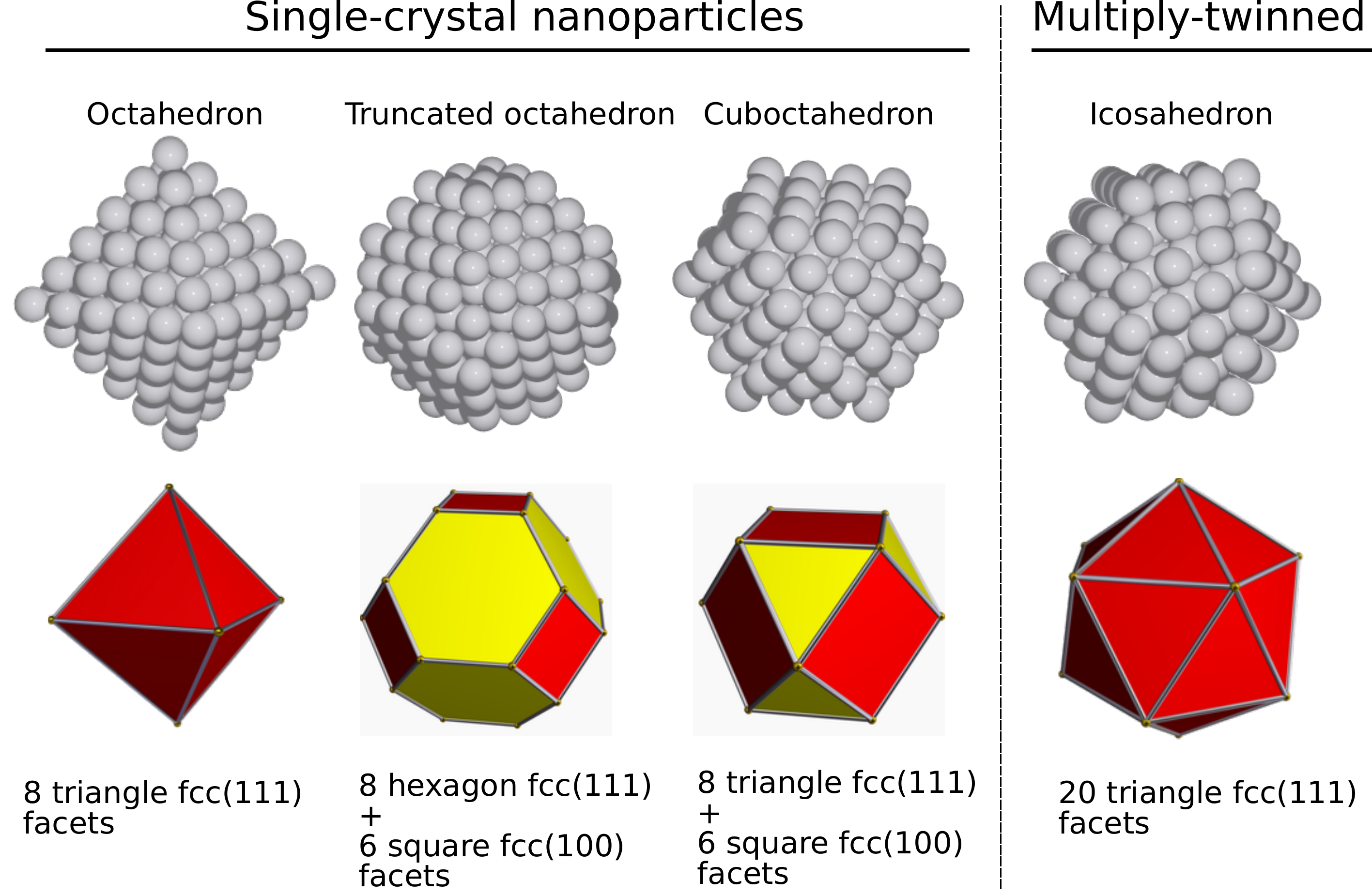}
\caption{Atomic configurations and geometrical shapes of nanoparticles in varioius shapes.}
\label{fig:np_diagram}
\end{figure}

Pure Pt nanoparticles are normally stable in the single crystal structure, whereas Pt alloy nanoparticles can exist in multiply-twinned structures such as an icosahedron~\cite{He2016, Xia2009, Wu2012, Shao2011}, although the shape of Pt nanoparticles can be controlled by capping materials~\cite{Ahmadi1996}.
It is thus crucial to understand the characteristics of structure motifs, such as icosahedron and octahedron.
An icosahedron is created by packing twenty tetrahedra in a matter that they share a common vertex, leading to close-packed surfaces, but distorted tetrahedra.
(See Figure~\ref{fig:np_diagram}.)
This distortion leads to a high internal strain with a relatively low surface energy~\cite{Mackay1962}.
Thus, icosahedra are normally more stable at small sizes, where the surface energy prevails over the volume contribution~\cite{Cleveland1991}.

In contrast, an octahedron structure preserves the bulk lattice symmetry---it can be obtained by directly cutting a single crystal.
The normal octahedron only has fcc(111) facets and has no internal strain.
To lower the total energy, the six ``tips'' of the octahedra can be removed, creating a truncated octahedron~\cite{Uppenbrink1992} (Figure~\ref{fig:np_diagram}).
Although this raises the energy per surface area by creating six fcc(100) facets, it lowers the total amount of surface area, which suggests it may be more stable at larger sizes~\cite{Baletto2002}.
The cuboctahedron can be conceptualized in similar way: starting with a cube with six fcc(100) facets, the eight corners are cut off, exposing eight new triangular fcc(111) facets (Figure~\ref{fig:np_diagram}).

The trade-off between surface and volume contributions lead to possible crossovers in stability among structural motifs as the particle size changes; for pure nanoparticles, past researchers have used simple empirical thermodynamic models to describe this phenomenon~\cite{Nishioka1977, Northby1989, Uppenbrink1992, Baletto2002}.
In the case of bimetallic nanoparticles, crossovers among various shapes have rarely been reported.
Zhu \etal\ used an empirical model fit to density functional theory (DFT) calculated properties to show that alloying of Pd can extend the stability of icosahedron Pd--Au nanoparticles beyond that of pure Au nanoparticles because of the stress release when two different metals are mixed~\cite{Zhu2015}.
Since the empirical potential was fit to a small number of properties, the prediction accuracy over a large range of different nanoparticle structures is uncertain.

The crossovers in bimetallic nanoparticles are not likely to be fully understood without reliable descriptions of the atomic interactions.
In an ideal world, such studies would employ electronic structure calculations directly.
DFT offers a good compromise between accuracy and computational cost.
However, model systems of practically-sized nanoparticles, with sizes of 1,000--10,000 atoms, are notoriously expensive for even single calculations, due to the famous $\mathcal{O}(N^{3})$ scaling, where $N$ indicates the system scale, such as the number of electrons, number of atoms or number of basis functions.
For nanoparticle structure exploration, the picture is more grim due to the huge combinatorial space that must be explored to describe the range of sizes, crystal structures, facets, and alloy (dis-)ordering that may be encountered.
Further, as a magnetic system, \CoPt\ particles have an added expense in electronic structure due to the requirement for spin-polarized calculations.
These combinations of factors make the rigorous exploration of \CoPt\ structure--function relationship out of reach for methods like DFT.

Machine-learned inter-atomic potentials have gained momentum in fitting potential energy surfaces calculated by \textit{ab initio} calculations~\cite{Behler2007, Bartok2010, Rupp2012,Behler2021,Deringer2021,Unke2021,Botu2017}.
However, a fundamental problem arises in using atomistic machine learning to perform large-scale simulations: for high-fidelity potentials, the training data should closely resemble the ultimate structures being predicted, and the generation of such training data for nanoparticles can be very costly, as described above.
To circumvent this issue, we recently released a ``nearsighted force-training'' (NFT)~\cite{Zeng2022} approach to generate small-size training data to systematically learn the forces and energies of large structures.
In this approach, a bootstrap ensemble~\cite{Peterson2017} (or any other reasonable uncertainty metric) is used to identify the most uncertain atoms in a particular structure.
Atomic ``chunks'' centered on these uncertain atoms are removed, and calculated at large enough size that the central atom's force can be calculated with fidelity by DFT.
Only the force on the central atom is used in the loss function, thus adding targeted data and avoiding noise associated with boundary atoms.
We demonstrated that this approach successfully built and relaxed nanoparticles containing up to 1,415 atoms in previous work.

Here we apply the NFT approach to build a robust ensemble model for Co--Pt nanoparticles.
Using these well-validated neural network models, we optimize the structures of Co--Pt alloys of simple bulk, fcc(111) surfaces, fcc(100) surfaces, icosahedron and octahedron nanoparticles of various sizes and compositions.
Moreover, we aim to address a number of key problems regarding the thermodynamic stability of Co--Pt nanoparticles, including the crossover among Pt and Co--Pt structure motifs, the most stable atomic arrangement of a Co--Pt truncated octahedron, and the order--disorder phase transition of Co--Pt truncated octahedron.

\section{Methods}\label{sec:ch-nft-ga-method}

\subsection{Model creation with nearsighted force training}

We used the nearsighted-force training (NFT) approach~\cite{Zeng2022} to generate the machine-learning model without the need for large-sized training data, which is expensive to obtain.
In this method, we started with a small training set consisting of 18 \CoPt\ bulk structures calculated in density functional theory (DFT), sampled following the initialization procedure we previously described.\cite{Zeng2022}
We trained a bootstrap ensemble~\cite{Peterson2017} of Behler--Parinello-type machine-learning models to this minimal training set; details of the model parameters are described later.
When this ensemble model is applied to a new structure, it gives a systematic estimate of the uncertainty on each atom in that structure.

We started the NFT active-learning procedure on a ``rattled'' \ce{Pt192Co68} cuboctahedron.
That is, we used the ensemble to identify the most uncertain atom in the structure, then we pulled out a ``chunk'' of this atom and its neighbors within a cutoff distance of 8~\AA.
This chunk was calculated in DFT, and solely the force on the central atom was added to the training set.
The ensemble was re-trained, and the procedure was repeated on the next most uncertain atom.
We terminated this process after 9 NFT steps, because the uncertainty did not improve in two consecutive steps, and we wanted our model to be optimized for low-force structures.

Next we performed a relaxation on this structure, and used the NFT model to extract uncertain chunks along the relaxation trajectory, with the procedure we described earlier~\cite{Zeng2022}, leading to a systematic improvement of the description of this relaxed particle.
We generally stopped the active learning process when any of three criteria were met: the uncertainty was below the convergence criterion, the number of retraining steps exceeded a pre-defined number, or the target uncertainty was not lowered for two continuous steps.

After we completed the NFT procedure on the \ce{Pt192Co68} particle, we continued to improve our model by training on a variety of octahedron and icosahedron nanoparticles, allowing us to find atomic chunks representing diverse local chemical environments that are unique and informative for potential energy surfaces of Co--Pt nanoparticles.
The icosahedron nanoparticles include \ce{Pt1415}, fully disordered \ce{Pt736Co679}, CoPt/2Pt core-shell  with a disordered core (\ce{Pt1150Co265}), CoPt/1Pt core-shell  with a disordered core (\ce{Pt3092Co1991}), and CoPt/2Pt core-shell with a disordered core (\ce{Pt3694Co1389}).
The cuboctahedron nanoparticles include \ce{Pt192Co68} with a disordered core, a \ce{Pt1415}, a core-shell CoPt/Pt cluster \ce{Pt975Co440} with a disordered core, a fully L1$_0$ ordered Co--Pt cluster \ce{Pt736Co679} and a fully disordered \CoPt\ cluster \ce{Pt736Co679}.
In total, 2064  atomic ``chunks'' in sizes from 38 to 168 atoms were extracted from those nanoparticles.
Adding the 18 bulk cells, we had in total 2082 training images.

The statistics of these 2064 generated atomic chunks are discussed in the SI, including the distributions of forces on the central atom, number of atoms, force prediction residuals and energy prediction residuals.
The nanoparticle trajectories, training images (atomic chunks), force and energy ensemble models are included as supporting datasets.

\subsection{Model structure}

We took a bootstrap approach~\cite{Peterson2017} to sample the training images for each neural network model, our ensemble consisted of 10 neural network models.
Ensemble averages were used as the predictions for both energy and forces.
Atomic uncertainties were proportional to standard deviations of the ensemble force predictions, as used in Zeng \etal. \cite{Zeng2022}
Each member of neural network models is a Behler-Parrinello type. \cite{Behler2007}
Neural network models were constructed using an open-source atomistic machine learning package (AMP) developed in our group. \cite{Khorshidi2016}
Fast force and energy inferences were carried out with n2p2, which is mainly developed by Singraber \etal.\cite{Singraber2019}
Fast fingerprinting was performed with SIMPLE-NN implemented with AMPTorch. \cite{Lee2019, Shuaibi2020}
Gaussian symmetry functions with a cutoff radius of 6.5 \AA~were used to encode the local chemical environments.
For the feature vector of Co elements, it consists of 12 G2 and 8 G4 symmetry functions, where Pt feature vector comprises 13 G2 and 7 G4 symmetry functions.
Full details of the symmetry functions are included in the supporting data set in the JSON format.
3000 epochs were used for training.
A simple structure of (20, 5, 5, 1) was employed for the neural network topology to mitigate overfitting.
A L$_2$ regularization (L$_2 = 0.001$) was used to avoid large atomic neural network weights, hence alleviating the overfitting.
We trained both a force ensemble model and an energy ensemble model on our 2082 images, referred to hereafter as the force model and energy model, respectively.
This was done in order to allow our model sizes to be smaller and our training procedures to be faster, with each individual model optimized for its own prediction of interest.
The force model was trained on forces of central atoms of atomic chunks and on both energy and forces of bulk cells.
The energy model was trained on the total energies of the bulk cells and atomic chunks.

\subsection{Electronic structure calculations}

DFT calculations for bulk, atomic chunks and 201-atom nanoparticles were carried out with the GPAW code~\cite{Enkovaara2010}.
The Perdew--Burke--Ernzerhof (PBE) exchange--correlation functional with a plane wave cutoff of 350 eV was used~\cite{Perdew1996}.
To achieve a fast convergence, a Fermi-Dirac smearing of 0.1 eV was utilized and the energetics were extrapolated to 0 K.
Calculations for atomic chunks were sampled at the $\Gamma$-point of the Brillouin zone, where calculations for bulk cells used a k-point grid of 12$\times$12$\times$12.
For atomic structures including cobalt, spin polarization was included.
When only platinum was present, the calculations were spin-paired.
The lattice constant of bulk Pt was found to be 3.936 \AA.
Atomic chunks were placed in a non-periodic box where the shortest distance to the box wall is at least 5 \AA.
Self-consistent field (SCF) calculations were considered to be converged when the energy difference between the last three steps is less than 0.0001 eV/electron.
Structure optimizations used a MDMin algorithm until the maximum atomic force was not larger than 0.05 eV/\AA.

DFT calculations of the 586-atom particles were done using the highly parallel ``Simulation Package for Ab-initio Real-space Calculations'' (SPARC) code.\cite{Xu2021, Ghosh2017, Ghosh2017b}
To the best of our knowledge, this calculation represents the largest size of Co--Pt nanoparticles that has ever been studied directly with DFT.
A mesh spacing of 0.13 \AA\ (0.25 Bohr radii) was used in a $\Gamma$-point calculation with the PBE functional and the PseudoDojo pseudopotentials~\cite{Setten2018}, and all calculations were run until the energy converged to within $2.7\times10^{-4}$ eV/atom ($1\times10^{-5}$ Ha/atom).
Atomic forces were computed and compared with GPAW results for smaller systems, which showed the mean absolute force error between the codes is below 0.025 eV/\AA~ for all systems tested (see Figure S3 in the SI).
Particles were surrounded by 3.5 Angstrom of vacuum in each direction with Dirichlet boundary conditions in all directions.

Note that unlike our prior publication~\cite{Zeng2022}, in this work we dealt with a magnetic system.
The underlying assumption for NFT to be applicable for magnetic systems is that in the electronic ground state there exists a unique mapping from atomic positions to spin states (or magnetic moments) which also display a strong locality.
In general, forces may depend on the initial guess of magnetic moments, since a poor initial guess may lead to a different local-minimum spin configuration.
Thus, we have used a consistent initial-guess strategy for all calculations, with an initial magnetic moment of 0 $\mu_\mathrm{B}$ for Pt atoms and 2.1 $\mu_\mathrm{B}$ for Co atoms.

\subsection{Global optimization techniques}
We aimed to explore a wide range of the potential energy surface with an emphasis on structures of nanoparticles in various shapes.
We were particularly interested in the global minima for a given shape, size and alloy composition.
However, searching global minima using a brute-force approach is computationally prohibitive.
For example, if we consider a small fixed-shape 147-atom nanoparticle with 73 Co and 74 Pt atoms, the number of possible atomic arrangement is already as large as $ 147!/(73! \times 74!) \approx 10^{44}$.
Although symmetry can reduce the complexity, attempting to exhaust the search space is inaccessible even with machine-learned potentials, especially for large nanoparticles in the size of thousands of atoms.
Instead, we used global optimization techniques; specifically genetic algorithms and Metropolis Monte-Carlo simulations.

\subsubsection{Genetic algorithms}

Genetic algorithms, inspired from evolutionary theory, have become popular in the past two decades for optimizing structures~\cite{Deaven1995, Johnston2003,Vilhelmsen2012, Lysgaard2014, Bossche2018}.
The genetic algorithm was performed on both Co--Pt surfaces and a type of Co--Pt icosahedron to construct energy convex hulls.
The genetic algorithm was set up with the Atomic Simulation Environment (ASE)~\cite{HjorthLarsen2017} based on the procedure implemented by Lysgaard \etal~\cite{Lysgaard2014} and Bossche \etal~\cite{Bossche2018}.
We used a rigid structure; that is, all derived structures are not allowed to relax.
Structural relaxation was performed afterwards if needed.
To make comparisons between different compositions, we define the negation of the mixing energy $E_\mathrm{f}(\mathrm{Pt}_x\mathrm{Co}_y)$ of a structure $\mathrm{Pt}_x\mathrm{Co}_y$ as the fitness score to propagate the algorithm:

\begin{equation}\label{eq:ga-fitness}
E_\mathrm{f}(\mathrm{Pt}_x\mathrm{Co}_y) = {E}(\mathrm{Pt}_x\mathrm{Co}_y)
- \frac{x}{x+y} {E}(\mathrm{Pt}_{x+y})
- \frac{y}{x+y} {E}(\mathrm{Co}_{x+y})
\end{equation}

\noindent
where $E(\mathrm{Pt}_x\mathrm{Co}_y)$, $E(\mathrm{Pt}_{x+y})$, and $E(\mathrm{Co}_{x+y})$ denote the ML model calculated per-atom energies of the corresponding structures.
We first studied fcc(111) and fcc(100) surfaces.
For the fcc(111) and fcc(100) surfaces, we used a $4\times4\times5$ supercell in size of 80 atoms with 13 \AA~ separation between slabs in the direction orthogonal to the surface.
The initial generation was populated with 120 surfaces using randomly chosen compositions.
The lattice constants of pure Co and Pt slabs were determined by the force model, and for the mixed slabs, the lattice constant is linearly interpolated based on Vegard's law.
We used three ASE operators to create the next generation.
``CutSpliceCrossover'', as introduced by Deaven and Ho~\cite{Deaven1995}, takes two parent structures, then cuts them in a random plane and combine the halves from two parent slabs together to form an offspring.
The second operator ``RandomSlabPermutation'' was used to randomly permute two atoms of different types.
The last ``RandomCompositionMutation'' changes the composition of the slab by mutating one element to the other.
The probability of the above three operators are respectively 0.6, 0.2, and 0.2.
In addition, we used a variable function named ``RankFitnessPopulation'' to uphold the composition diversity at each generation so that optimization is on a full range of compositions.
We ran for 100 generations.

This approach was also used to build the convex hull of a 147-atom \CoPt\ icosahedron nanoparticle.
The initial generation was populated with 100 members, the composition was randomly chosen, and the lattice constant was obtained by a linear interpolation.
For the icosahedron nanoparticle, we were interested in the fittest Co/Pt composition, hence we did not restrict the algorithm to keep a wide range of compositions in each generation.
For the nanoparticles, four types of operations were utilized to create the offspring, including ``CutSpliceCrossover'', ``RandomSlabPermutation'', ``MirrorMutation'' (to mirror half of the cluster in a randomly oriented cutting plane while discarding the other half) and ``SymmetricSubstitue'' (to permute all atoms within a shell of the symmetric particle), and the corresponding operation probabilities are 3/6, 1/6, 1/6, 1/6, respectively.
This was also run for 100 generations.
After the runs were completed, we chose to study structures after the 80th generation.
We pinpointed the Co--Pt icosahedron with the most negative formation energy, which is named as the fittest Co--Pt icosahedron.
We calculated the $c_{\ce{Pt}}$ of the fittest structure, and we selected structures whose mole fractions of Pt are close to the fittest one based on a Gaussian function.
In total, 232 Co--Pt icosahedron nanoparticles were selected, and energetics of their relaxed structures were used to construct the energy convex hull for the 147-atom Co--Pt icosahedra.

\subsection{Metropolis Monte Carlo simulations}

Previous works suggested that while a genetic algorithm was more efficient to search a wide range of compositions, Metropolis Monte Carlo simulations were found to be more effective for structures with fixed compositions~\cite{Boes2017, Muller2005, Yang2006}.
We thus employed such Metropolis calculations, in the canonical ensemble.
At each elementary Monte Carlo step, two neighboring atoms were exchanged and the energy change was calculated for the exchange.
The new structure was accepted if the energy change was negative, or it is accepted based on the Boltzmann probability if the energy change was positive.
The number of Monte Carlo steps was determined in a way that on average at least 40 swaps were performed for each atom in the structure.
We used these simulations at 300 K to find the putative global minima of truncated octahedra \ce{Pt96Co105} and \ce{Pt300Co286}, and we compared them to the fully L1$_0$ ordered counterpart at full DFT levels of theory.
We also used these simulations at temperatures ranging from 300 K to 1800 K to study the order--disorder phase transitions for Co--Pt bulk and nanoparticles with nearly equal compositions of Co and Pt.
For each Monte Carlo trajectory at a given temperature, the order parameter was calculated as the average over configurations after a burn-in period.

\section{Results \& Discussion}

The key objective of this study is to train robust machine learning models that can predict stable structures of nanoparticles, which we will use to distinguish the phase stability of bimetallic nanoparticles with various shapes and atomic arrangements.

\subsection{Comparison to literature structures and \textit{ab initio} calculations}

We first validated our ML models by a comparison with published \textit{ab initio} calculations; additional validation calculations are reported in Sections~\ref{sec:ga} and \ref{sec:order-disorder} and comparisons to literature-reported structures and trends are contained in Sections~\ref{sec:ga}, \ref{sec:crossovers}, and \ref{sec:order-disorder}.

\begin{figure}
\centering
\includegraphics[width=4.in]{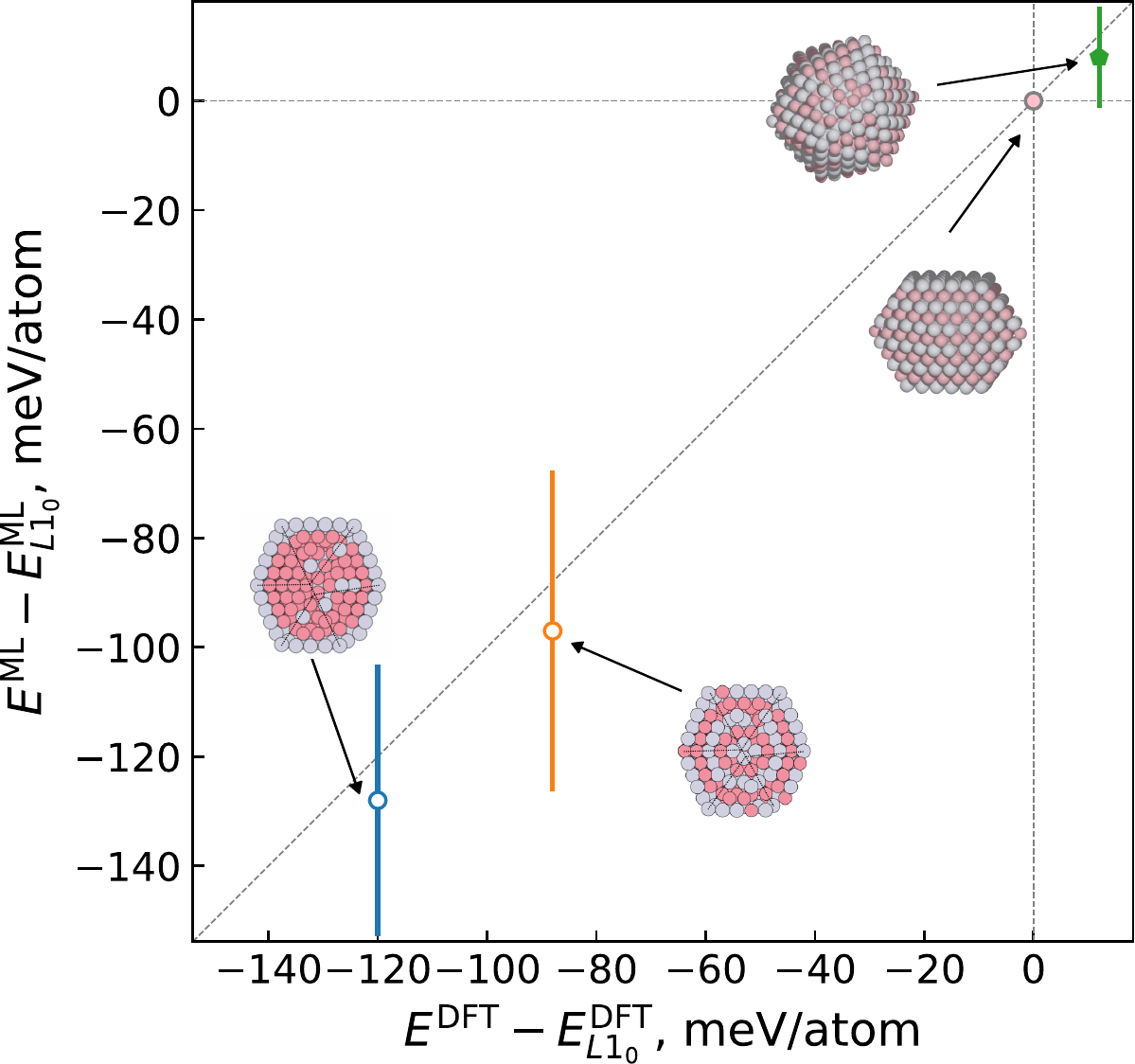}
\caption{Energetics of \ce{Pt296Co265} nanoparticles in various shapes in reference to the L1$_0$ ordered cubocahedron: ML predictions versus DFT calculations by Gruner \etal~\cite{Gruner2008}. The DFT calculations used a smaller cutoff (268 eV) compared to 350 eV used to obtain the training data for ML models. The error bar represents the ensemble halfspread as defined by Peterson et al~\cite{Peterson2017}.
\label{fig:shape_validation}}
\end{figure}

Gruner \etal~used first-principles calculations to compare the energetics of several structural motifs of Co--Pt alloy nanoparticles in reference to an L1$_0$ ordered cuboctahedron~\cite{Gruner2008}.
We created a number of 561-atom nanoparticles for \ce{Pt296Co265} that are either identical or close in atomic arrangements to those used in Gruner's \textit{ab initio} calculations, since the exact structure was not always reported.
Different randomness should play a negligible role in the energetics because only a small fraction of atoms are randomly positioned.
The nanoparticles included an L1$_0$ ordered cuboctahedron, a disordered icosahedron, an icosahedron with alternating Co and Pt shells, and a core--shell icosahedron with a Co-rich second shell.
We relaxed these structures with the force model and then we calculated the energetics with the energy model.
The comparison between ML predictions and \textit{ab initio} calculations by Gruner \etal\ is shown in Figure~\ref{fig:shape_validation}.
We note that the work of Gruner \etal\ used a smaller cutoff (268~eV) than the one (350~eV) we used for DFT calculations on atomic chunks, which may account for some discrepancy to the literature.
One can see a very good agreement for both cuboctahedron and icosahedron nanoparticles, and the overall order for all structures presented is exactly captured by the ML models, with the mean ensemble prediction agreeing very well and the parity line within the error bars.

In addition, we created a 147-atom Pt icosahedron and cuboctahedron, relaxed it with the ML model, and compared the absolute energy difference to that obtained by DFT calculations we performed in the GPAW calculator.
The DFT and ML-predicted energies for both structures are presented in Figure S4 of SI.
Although the exact energetics for each shape can differ by 13.6--27.2 meV/atom between ML predictions and DFT calculations, the relative energy difference between those two shapes is much closer; Pt icosahedron is more stable than Pt cuboctahedron by 7 meV/atom using ML models versus 8.6 meV/atom using DFT calculations.
This suggests that the ML models are able to distinguish the thermodynamic stability across various shapes of nanoparticles, and different atomic arrangements for a given shape.

\subsection{Energy convex hull of Co--Pt surfaces and nanoparticles}\label{sec:ga}

We next turn to studying the relative stability of alloyed structures, by using these potentials to construct ``convex hulls'' that plot the alloy-formation energy versus composition.
Negative energies indicate the alloy is stable relative to the pure components.

We began by examining the formation energy of two ordered bulk alloys, \ce{PtCo} and \ce{Pt3Co}.
The formation energy for  \ce{PtCo} and \ce{Pt3Co} are -0.24 and -0.14 eV/atom, respectively, close to values by experiments and empirical potentials~\cite{Oriani1962, Andreazza2015}.
As a comparison, the DFT-calculated formation energies for \ce{PtCo} and \ce{Pt3Co} are -0.10 and -0.06 eV/atom, respectively.
This indicates that Co and Pt atoms have a strong tendency of being mixed.

\begin{figure}
\centering
\includegraphics[width=6.0in]{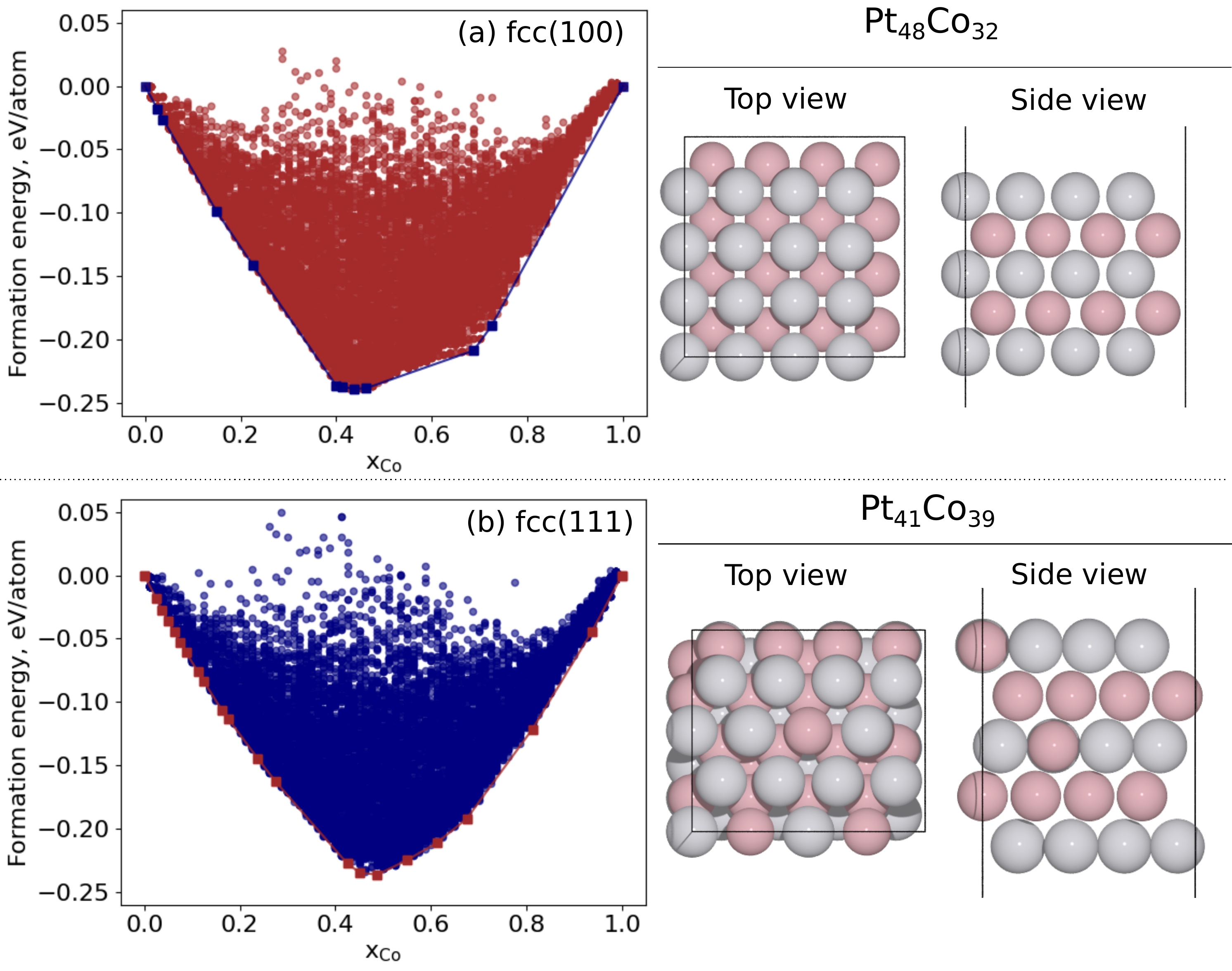}
\caption{Energy convex hulls of fcc(100) (a) and fcc(111) (b) surface PtCo alloys as a function of Co compositions. Solid squares are the stable structures and circles refer to unstable structures. Top and side views of the fittest configurations are shown on the right.}
\label{fig:convex-hull-surface}
\end{figure}

To gain insights into the atomic arrangement near a Co--Pt surface, we built energy convex hulls for 5-layer Co--Pt fcc(100) and fcc(111) surfaces using the neural-network-enhanced genetic algorithm.
Figure~\ref{fig:convex-hull-surface}(a) shows the energy convex hull for fcc(100), where $x_{\rm{Co}}$ represents the mole fraction of Co atoms.
We extracted the fittest fcc(100) surface (that with the lowest formation energy), and its composition is around $x_{\rm{Co}}=0.4$.
A side view of the global minima, shown in the figure, implies that it is an L1$_0$ ordered structure, forming alternating Pt and Co layers with the outermost layer being Pt.
Although the convex hull of fcc(100) is not symmetric, we can infer that this lack of symmetry is constrained by the number of layers (5), and if it were increased it would approach a more symmetric form.
The structure at $x_{\rm{Co}}=0.2$ is close to L1$_2$ ordered, which suggests the atomic arrangement in a fcc(100) surface is inclined to form an ordered structure.
We anticipate that compositions of stable structures may change if we increase the thickness of the surface where bulk contributions become more dominant (\textit{e.g.}, global minima closer to 0.5 for an infinitely thick surface).
Yet atomic arrangement patterns, in particular near surfaces, should hold for thicker surfaces.
We will see this to be the case when we examine large nanoparticles.

In comparison, the convex hull of fcc(111) surface alloys is much smoother, representing more flexibility in atomic arrangements in this more closely-packed facet.
In terms of the fittest configuration \ce{Pt41Co39}, the general trend still holds that Pt tends to segregate at the surface while being depleted at the subsurface.
A small amount of Co shows up at the surface, although Co--Co direct connections are not present.
In brief, the main features of atomic arrangements for both fcc(100) and fcc(111) surfaces are a surface Pt layer and a subsurface Co layer.
Besides, the fcc(100) surface is more likely to form an ordered structure than the fcc(111) surface.

\begin{figure}
\centering
\includegraphics[width=4.5in]{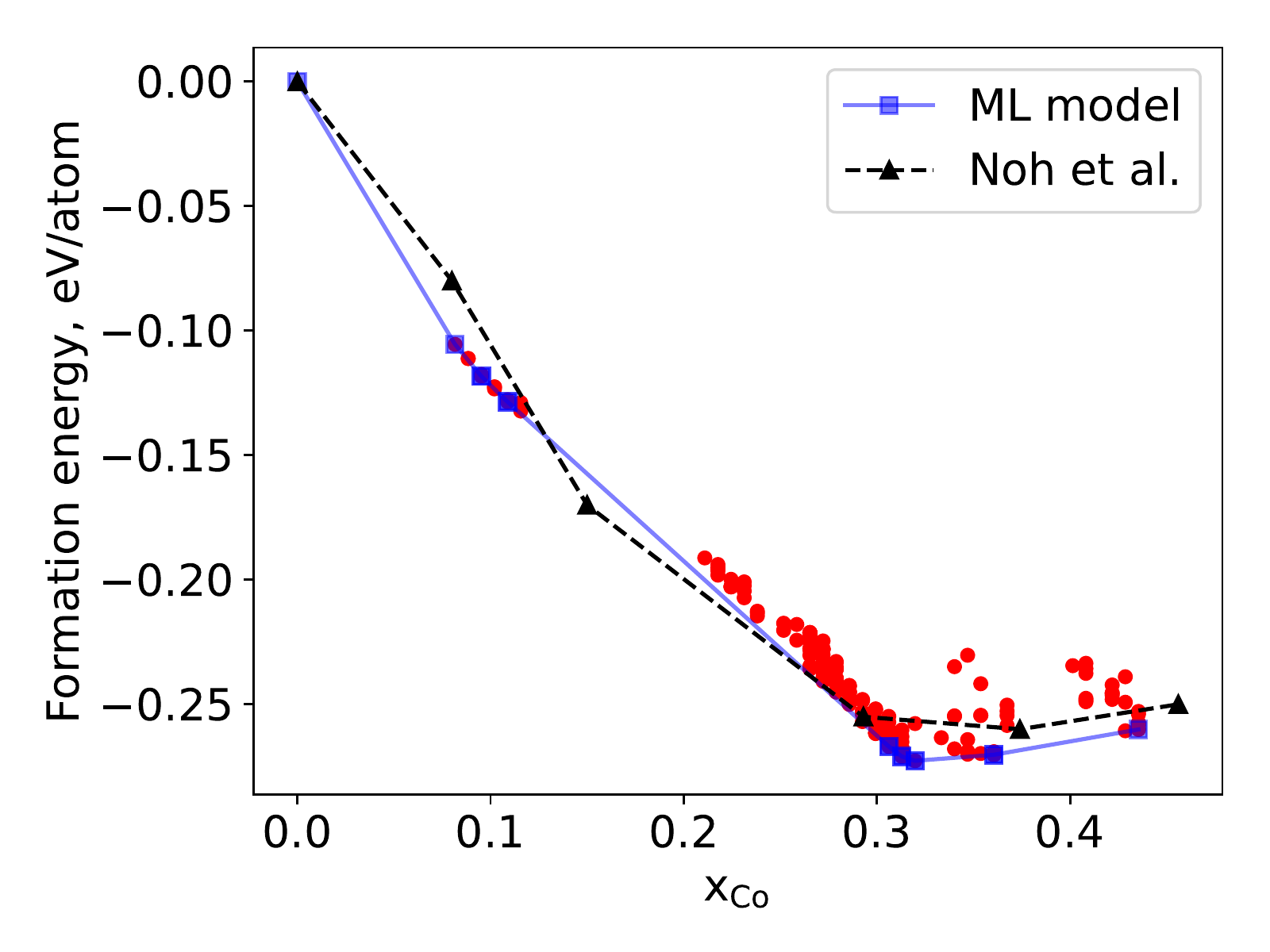}
\caption{Energy convex hull of a 147-atom Co--Pt Ih as a function of Co compositions. Solid squares are the stable structures found by ML models, and circles refer to unstable structures of ML predictions. Up-triangles refer to the stable structures excerpted from Noh \etal~\cite{Noh2013}.}
\label{fig:convex-hull-Ih}
\end{figure}

Next we turned to build the convex hull of a 147-atom Co--Pt icosahedron.
Figure~\ref{fig:convex-hull-Ih} shows the formation energy for this system as a function of mole fractions of Co atoms, centered on the Pt-rich region where the global minimum lies.
The convex hull from a DFT-based study conducted by Noh \etal~\cite{Noh2013} is also included for comparison.
The trend of formation energy versus composition demonstrates that the ML-predicted convex hull agrees very well with the \textit{ab initio} results.
The prediction discrepancy of formation energies is less than 0.03 eV/atom across the composition range shown in Figure~\ref{fig:convex-hull-Ih}.
The discrepancy could likely be reduced by adding additional atomic chunks extracted from uncertain 147-atom Co--Pt icosahedra---here, we did not seek such an improvement since we considered the prediction accuracy to be satisfactory.
On the Pt-rich side (increasing from $x_{\ce{Co}}=0$), the formation energy rapidly drops with the addition of Co.
In contrast, there exists a wide flat region ($0.3 \leq x_{\rm{Co}} \leq 0.5$) where either addition or removal of Co atoms barely changes the formation energy.

The most significant feature (also reported by Noh \etal) along the convex hull is the formation of a Pt skin on the surface with a Co-rich layer directly below the Pt skin.
Among the eight Co--Pt icosahedra along the hull in Figure~\ref{fig:convex-hull-Ih}, seven structures are covered by a full Pt skin, with the exception being the one with the lowest overall platinum composition ($x_{\ce{Co}}=0.44$), where the skin contains 78\%\ Pt.
The average Co composition of the second shell is 74\%.

It is well-known in fuel-cell catalysis that Pt--Co alloyed catalysts exhibit a platinum skin, which is generally considered to be formed by dealloying of the cobalt under electrochemical conditions, due to the difference in electrodeposition potential between Co and Pt.
The Pt skin is contracted relative to what would be found on a pure-Pt particle, which has been concluded to increase the catalyst's activity.
These results show that Pt is also thermodynamically most stable on the surface (for a fixed Pt:Co composition).
This may suggest a greater long-term stability to these catalysts than if the skin layer were present due to cobalt dissolution alone.

On the left side of the convex hull, the ML model identifies a stable structure with composition \ce{Pt83Co64}, where Co atoms on the surface occupy the center of fcc(111) surfaces.
In this same region, Noh \etal's calculations show a structure with composition \ce{Pt80Co67}, where Co atoms on the surface sit at the corners.
To validate whether the center occupancy represents a stable atomic arrangement, we constructed icosahedron structures with stoichiometry \ce{Pt80Co67} where surface Co atoms occupy both types of sites, and compared the energies with both DFT and ML calculators.
In both cases, the corner occupancy was predicted to be more favorable, with DFT energies  showing a difference of $\sim$14 meV/atom, which indicates that the terrace center occupancy is a low-energy state as well.
However, this configuration was not captured in the DFT calculations by Noh \etal~\cite{Noh2013}.
The configurations of two types of \ce{Pt80Co67}, together with their ML and DFT energies,  are provided in Figure S5 of SI.
Since both calculators correctly show the corner site to have lower energy, this indicates that the two procedures captured different minima structures purely by the stochasticity of the genetic algorithm itself, and not due to an issue with the ML fidelity.

\subsection{Crossovers among morphology in Pt and Co--Pt nanoparticles\label{sec:crossovers}}

\paragraph{Platinum particles.}
In this section, we aim to provide physical insights into the distinct crossover behavior of Co--Pt nanoparticles.
We first focus on pure Pt nanoparticles, where the crossover between different shapes has been extensively investigated based on well-parameterized empirical potentials~\cite{Uppenbrink1992, Baletto2002, Baletto2005}.
The energies of each structure in such studies were fit to an empirical thermodynamic model, dividing the total potential energy ($U$) for a nanoparticle of a specific shape into contributions from volume, surface, and edges:

\begin{equation}\label{eq:crossover-model}
\frac{U}{N} = A + B N^{-1/3} + C N^{-2/3}
\end{equation}

\noindent
where $N$ is the total number of atoms, and $A$, $B$, and $C$ are parameters corresponding to the volume, surface, and edge contributions, respectively.
These parameters are unique to each nanoparticle shape.
As $N$ increases, the edge contribution becomes less important, and we will show that this term can be dropped in our size range of interest.

Discrepancy exists in the literature, even for the crossover of pure Pt nanoparticles.
For example, Uppenbrink \etal concluded that the crossover between icosahedron and cuboctahedron occurs at around 550 atoms for both pure Pt and pure Pt nanoparticles, and a decahedron is found in a narrow range of sizes~\cite{Uppenbrink1992}.
In contrast, Baletto \etal, using a different empirical potential, concluded that the crossover between icosahedron and regular truncated octahedron should occur in a small size ($<$100 atoms) and decahedra can exist in a wider range of sizes~\cite{Baletto2002}.
For simulations on small-size Pt nanoparticles, either cuboctahedron or regular truncated octahedron has been used in previous works~\cite{Tang2010, Shao2011, Alloyeau2009}.
Although controlling experimental conditions can open up possibilities for a variety of shapes of pure Pt nanoparticles, it is well acknowledged that multiply-twinned structures rarely form~\cite{Xia2009}.

We used the well-validated ML models to predict energetics of typical structure motifs of Pt nanoparticles across a size range of 201 to 6266 atoms, including seven cuboctahedron, five truncated octahedron, and seven icosahedron nanoparticles; decahedron is not considered because it is usually only an intermediate state and it has been rarely reported in experiments~\cite{Xia2009, Wang2015}.
We then fit the predicted energetics of each structure type as a function of $N$ to a simplified version of equation~\eqref{eq:crossover-model} in which we dropped the edge term ($C N^{-2/3}$).

\begin{figure}
\centering
\includegraphics[width=5.0in]{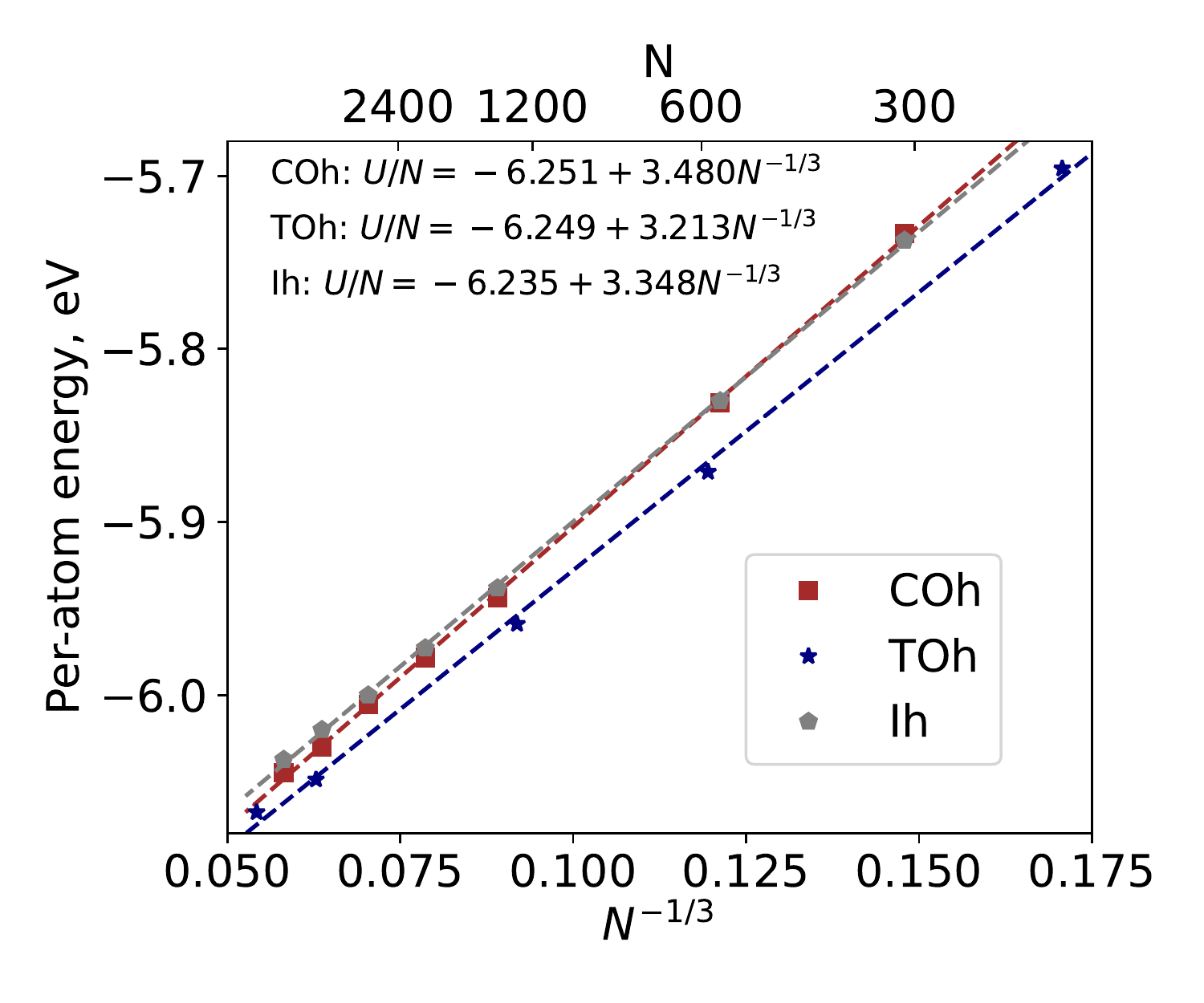}
\caption{
Energies of relaxed structure motifs of Pt nanoparticles, plotted as per-atom energy ($U/N$) versus $N^{-1/3}$.
COh, TOh and Ih represent cuboctahedron, truncated octahedron and icosahedron, respectively.
\label{fig:pt-crossover}}
\end{figure}

The results are shown in Figure~\ref{fig:pt-crossover}, which shows per-atom energy versus $N^{-1/3}$.
(Results including edge terms are included in Figure S6 of SI.)
First, we note that the data points show little significant curvature about the straight lines, which implies that the neglect of edge terms is justified for this range of particle size.
From the fit parameters displayed on the plot, one can see that volume contributions of single-crystal cuboctahedron and truncated octahedron are almost identical, while that of icosahedron is larger.
We attribute this to the distorted internal structure of the icosahedron.
As we expect, the surface contribution is always positive, with the order of cuboctahedron $>$ icosahedron $>$ truncated octahedron.
It is reasonable that the surface contribution of cuboctahedron is larger than that of icosahedron because more fcc(100) facets are exposed on the cuboctahedron surface compared to all fcc(111) facets on the icosahedron surface.
It was also found that the surface contribution of truncated octahedron is lower than icosahedron, probably because the distorted internal structure of icosahedron also has a profound impact on its surface energy (\textit{i.e.}, the icosahedron surface may be distorted as well).

This analysis shows the truncated octahedron to be the most stable Pt shape across this size range (roughly 200-7000 atoms).
To the best of our knowledge, it is the first time that the unique high stability of truncated octahedron is identified,  and it explains the observation that truncated octahedron is the structure of pure Pt nanoparticles most frequently found in experiments. \cite{Shao2011, Xia2009}
A crossover exists between cuboctahedron and icosahedron, which is estimated to be at $N=538$, in agreement with the result of Uppenbrink \etal. \cite{Uppenbrink1992}
However, these lines are nearly overlapping which presumably makes the precise location of the crossover very sensitive to the fidelity of the interatomic potential used, perhaps explaining the large variation seen in the literature for the location of this crossover.

\paragraph{Bimetallic particles.}
Using a similar logic, we extended this analysis to the crossover in Co--Pt nanoparticles.
Section~\ref{sec:ga} concluded that the most outstanding feature for Co--Pt nanoparticles is an outermost Pt skin with a Co-rich second shell.
Thus for truncated octahedron and icosahedron, we created Co--Pt structure motifs enforcing the above feature, with the remaining Co atoms randomly placed in the core.
As a comparison, we also included a cuboctahedron with an L1$_0$-ordered core surrounded by a pure Pt skin.
To account for the Co composition effect on the energetics, we introduced a revised empirical model, as shown in the following equations.
In the interest of having fewer parameters, we assume edge sites can be neglected (as we justified earlier for pure Pt particles in this size range).

\newcommand{\xco}{\ensuremath{x_{\ce{Co}}}}

\begin{equation}\label{eq:ptco-crossover-model}
    \frac{U}{N} = A(\xco) + B(\xco) \cdot N^{-1/3}
\end{equation}
where
\begin{equation}\label{eq:a-alloy}
A(\xco) = x_{\ce{Co}} A_{\mathrm{Co}} + (1-x_{\ce{Co}}) A_{\rm{Pt}} +
 \alpha x_{\ce{Co}} (1-x_{\ce{Co}})
\end{equation}
and
\begin{equation}\label{eq:b-alloy}
B(\xco) = B_0 (1 - \kappa x_{\ce{Co}})
\end{equation}

\begin{figure}
\centering
\includegraphics[width=6.0in]{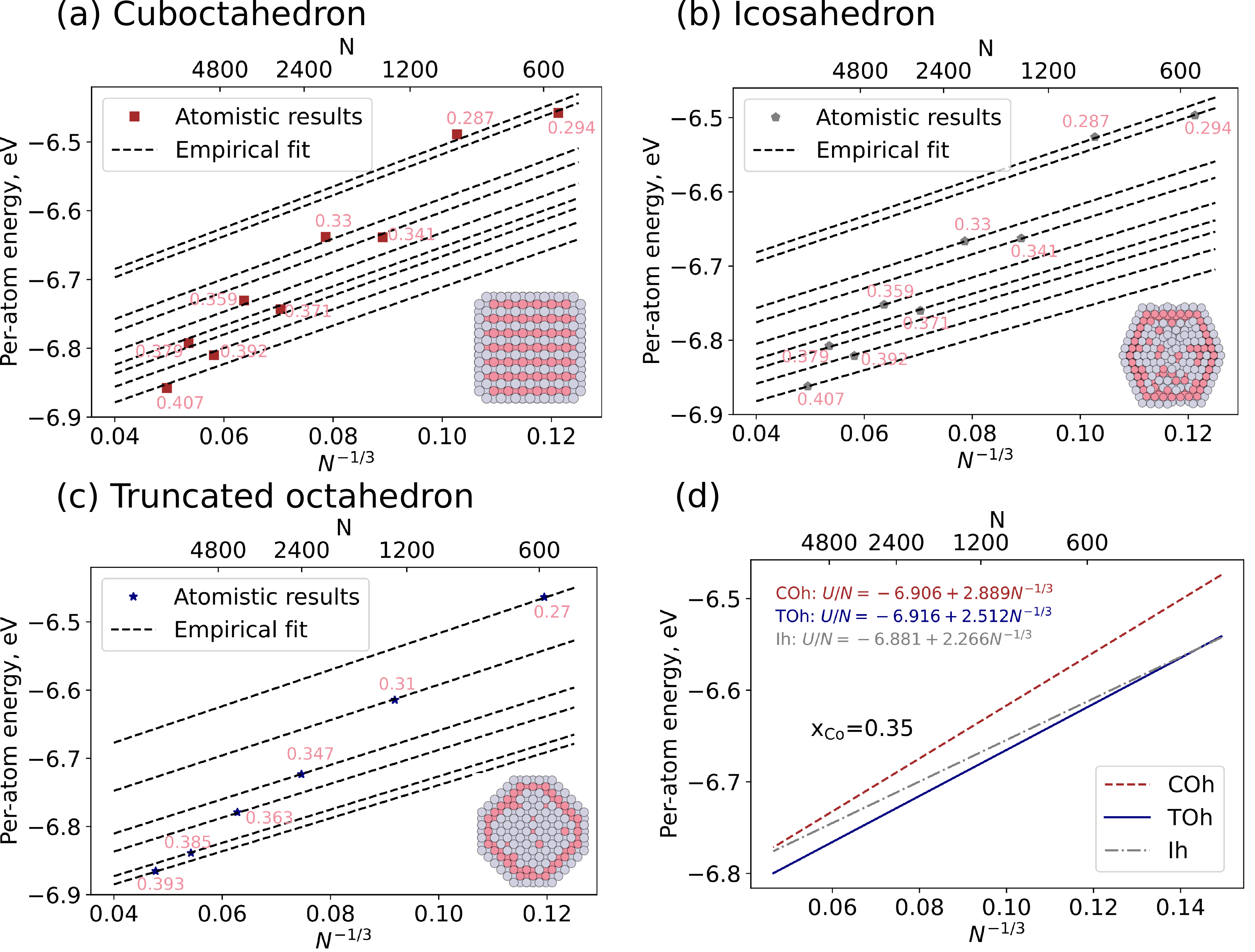}
\caption{Energies of relaxed structure motifs of \CoPt\ nanoparticles, plotted as per-atom energy ($U/N$) versus $N^{-1/3}$. COh, TOh and Ih represent cuboctahedron, truncated octahedron and icosahedron, respectively.
(a) Cubocahedron, (b) Icosahedron, (c) Truncated octahedron, (d) Fitted results using Eq.~\ref{eq:ptco-crossover-model}, at a Co composition of 35\%.
The Co compositions are indicated by texts next to each data point.
The fitted family of lines are indicated by dash lines; each line represents the fit for the fixed Co composition indicated by the nearest point.
Cross-sectional view of the structure motif for \CoPt\ nanoparticles are included as an inlet.
The number of atoms are presented in the twin axis at the top.
\label{fig:ptco-crossover}}
\end{figure}

\noindent
Here, $A_{\rm{Pt}}$ and $B_0$ are respective volume and surface contribution found from pure Pt nanoparticle results.
$A_{\mathrm{Co}}$ is the volume contribution in a pure Co nanoparticle; this was found from a bulk calculation.
$\alpha$ describes the enthalpy of mixing and $\kappa$ describes the reduction of the Pt surface energy due to alloying the bulk with Co; both of these were taken as free variables fit to the \CoPt\ particles in question.
The ML-calculated energetics for cuboctahedron, icosahedron, and truncated octahedron are shown in Figure~\ref{fig:ptco-crossover}(a--c), and the fitted results for each shape at each Co composition are represented by a family of dashed lines.
The fitted results are summarized in Figure~\ref{fig:ptco-crossover}(d) using a Co composition of 35\%, which falls within the range of investigated compositions.
Co alloying lowers both volume and surface contributions, suggesting a strong tendency of mixing Co and Pt atoms; however, the dependencies differ with Co--Pt nanoparticle shapes.
The fitted enthalpy of mixing and reduction in Pt surface contribution due to Co alloying for each shape of Co--Pt nanoparticle, together with fitted volume and surface contributions in a pure Pt nanoparticle and volume contribution of Co from a bulk calculation, are listed in Table~\ref{tab:fitted}.

\begin{table}
\centering
    \caption{Fitted enthalpy of mixing ($\alpha$) and reduction of Pt surface energy due to Co alloying ($\kappa$) for each shape of Co--Pt nanoparticles. Fitted volume ($A_{\rm{Pt}}$) and surface contributions ($B_0$) in a pure Pt nanoparticle are also listed. The volume contribution of Co ($A_{\mathrm{Co}}$) obtained from a bulk calculation is shown as well.
\label{tab:fitted}}
\begin{tabular}{lccccc}
\hline \hline
 & $\alpha$ & $\kappa$ & $A_{\ce{Pt}}$ & $B_0$ & $A_{\mathrm{Co}}$ \\
 & [eV/atom] & [eV/atom$^{4/3}$] &  [eV/atom] &  [eV/atom$^{4/3}$] & [eV/atom]\\
\hline
Cuboctahedron & -0.913 & 0.485 & -6.251 & 3.480 & -7.528\\
Icosahedron & -0.850 & 0.924 & -6.235 & 3.348 & -7.528\\
Truncated octahedron & -0.966 & 0.624 & -6.249 & 3.213 & -7.528\\
\hline \hline
\end{tabular}
\end{table}

Over most of the size range, the truncated octahedron is the most stable shape, as it is for pure Pt particles.
We can deduce, by extrapolation, that at small particle sizes ($<$333 atoms) icosahedron becomes more stable.
We infer that this is because the addition of Co in the core switches the order of the surface contribution coefficients for alloyed truncated octahedron (2.512) and icosahedron (2.266) compared to that for pure Pt truncated octahedron (3.213) and icosahedron (3.348), adding that the order of volume contributions remain unchanged for alloyed truncated octahedron (-6.916) and icosahedron (-6.881) versus that in pure Pt truncated octahedron (-6.249) and icosahedron (-6.235), as shown in Figure~\ref{fig:pt-crossover} and Figure~\ref{fig:ptco-crossover}(d).
This can be explained by the strain/stress release on the distorted surface of icosahedron when a smaller element such as Co is introduced into the subsurface and core.
Specifically, we define the average local strain for a surface atom $i$ as:
\begin{equation}\label{eq:local_strain}
\varepsilon_i = \frac{1}{M} \sum_{j \in \textrm{NN}_i} \frac{d_{ij} - d_{\rm{Pt}}}{d_{\rm{Pt}}}
\end{equation}
where $d_{ij}$ is the interatomic distance between atoms $i$ and $j$, $d_{\rm{Pt}}$ is the DFT-calculated lattice constant for an optimized bulk Pt (3.936 \AA), $\textrm{NN}_i$ represents the summation over the nearest surface neighbors of atom $i$, and $M$ is the number of surface neighbors.
We calculated the average local strains on the terrace fcc(111) sites for cuboctahedron and icosahedron particles in sizes from 561 to 5083 atoms with and without Co alloying, and report the result in Table~\ref{tab:strain}.
As we can see, the surface atoms on the cuboctahedron start in compressive strain for pure Pt, and alloying with Co only further compresses the surface atoms.
In contrast, the icosahedron starts in tensile strain in the pure system, so alloying with Co allows strain relief and crosses into the compressive regime.
As a result, the absolute strain increases with alloying for cuboctahedron and decreases for icosahedron, confirming stress release in icosahedron nanoparticles upon Co alloying.

If we increase the Co composition to 40\%, we found that the crossover between icosahedron and truncated octahedron shifts to a larger size of 570 atoms, further extending the range of stability for icosahedron.
Of course, the crossover may also depend on the surrounding environment and surface reactions, which are not considered in this study.
Here we main aim to providing the physical insights for differences in crossover for Pt and Co--Pt nanoparticles, and the structures used to analyze the crossover for Co--Pt alloy systems can probably be further optimized.
The method presented here can readily be extended.

\begin{table}
\centering
\caption{Strain levels (engineering strain, percent) for particles in sizes of 561, 923, 1415, 2057, 2869, 3871 and 5083 atoms with and without Co alloying. The $\pm$ indicates the standard deviation across particle sizes.
\label{tab:strain}}
\begin{tabular}{lrr}
\hline \hline
& Cuboctahedron & Icosahedron\\
\hline
Pure Pt & (-1.321 $\pm$ 0.176)\% & (+2.333$\pm$0.215)\% \\
Alloyed & (-3.116 $\pm$ 0.068)\% & (-1.244 $\pm$ 0.190)\% \\
\hline \hline
\end{tabular}
\end{table}

\subsection{Order--disorder phase transition in Co--Pt truncated octahedrons\label{sec:order-disorder}}

Before discussing the order--disorder phase transition, we need to first investigate the stable structure of a Co--Pt nanoparticle.
Two questions should be answered in this regard---first, is the stable structure ordered? Second, if it is ordered, how?
The truncated octahedron structure was chosen for this analysis based on the crossover analysis, as it is the most thermodynamically stable shape for large sizes and is also the structure most commonly reported in experiments~\cite{Xia2009, Li2019, Li2020}.
First, we performed Metropolis simulations at a temperature of 300~K on particles with composition \ce{Pt300Co286}.
We picked a structure after more than 58,600 steps, equivalent to 100 swaps per atom on average, and we treated it as the putative global minima.
We relaxed the structure using the force model.

\begin{figure}
\centering
\includegraphics[width=6.0in]{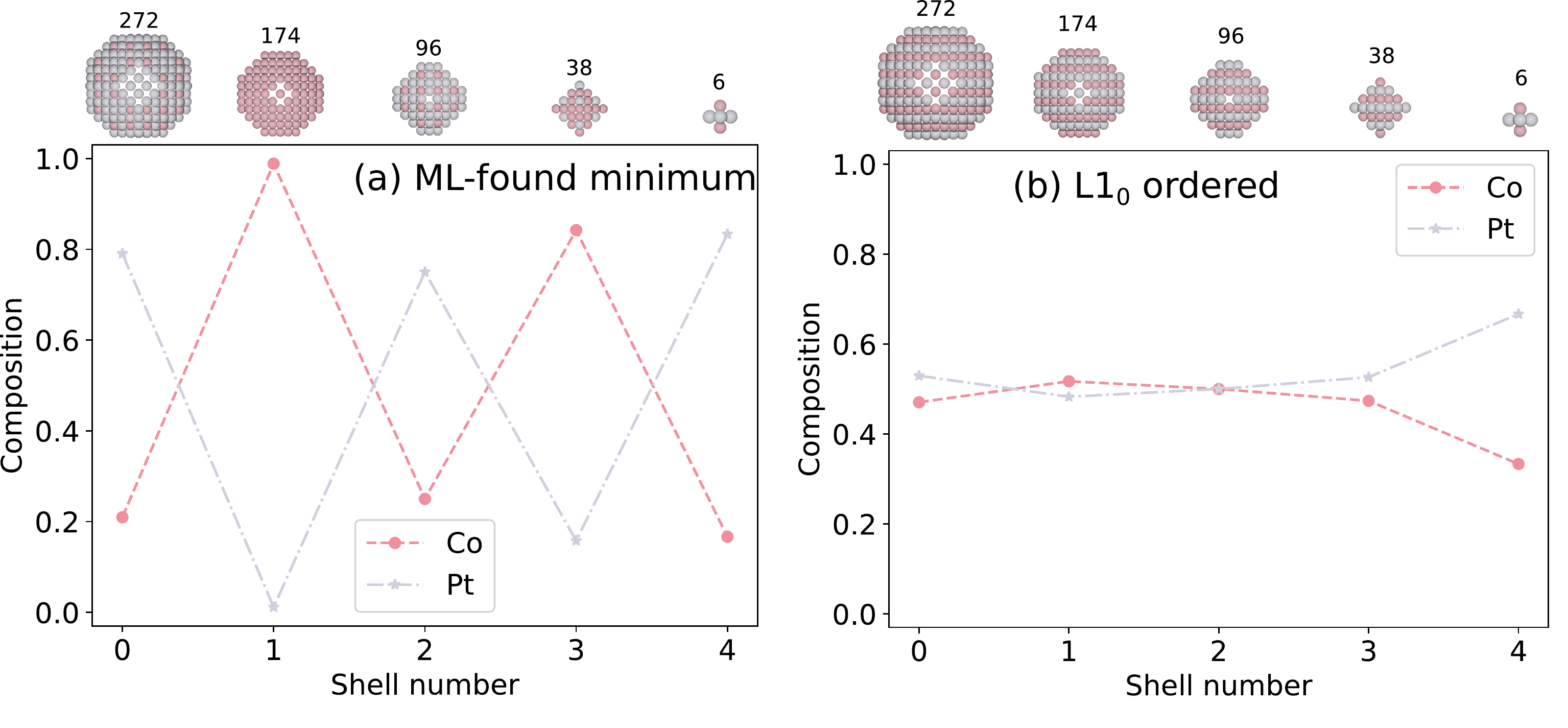}
	\caption{Composition depth profile of a truncated octahedron Pt$_{300}$Co$_{286}$:  the putative global minima found by ML models (a), and the fully L1$_0$ ordered Co--Pt nanoparticle alloy (b). Atomic arrangement at each shell and the total number of atoms are provided at the top.
\label{fig:gm-586}}
\end{figure}

We compared the energy of this structure with its fully-ordered L1$_0$ counterpart, using both our ML energy model and with DFT, using the highly-scalable SPARC code.
To our knowledge, this DFT validation calculation on a spin-polarized 586-atom structure represents the largest Co--Pt nanoparticle that has been directly validated by a full \textit{ab initio} method.
The shell-by-shell atomic arrangement of both structures are shown in Figure~\ref{fig:gm-586}.
For the ML-found minimum, an alternating preference for Pt and Co atoms can be seen starting with a Pt-rich surface with the subsurface layer fully occupied by Co.
Co atoms on the surface of this structure are more likely to occupy terrace fcc(111) sites and to connect with Pt atoms on the surface, consistent with the previous findings in the genetic algorithm study.
This observation is validated by aforementioned SPARC DFT calculations to prove that L1$_0$ ordered Co--Pt truncated octahedron alloy is not the most stable structure, but ordered in a different pattern as shown in the configurations in Figure~\ref{fig:gm-586}(a).
Both the SPARC DFT calculations and the ML-calculations conclude the Metropolis-derived structure found by ML models is lower in energy than the L1$_0$ structure, with ML predicting 0.097 eV/atom and DFT calculating 0.057 eV/atom.
The DFT-maximum atomic forces for the ML-found minimum and L1$_0$ ordered structures are 0.42 and 0.27 eV/\AA, respectively, which are within ML-predicted maximum atomic uncertainty of forces, 0.43 and 0.42 eV/\AA, respectively.
Since the Metropolis-found structure has a larger maximum force, we expect its energy may decrease more if it were re-optimized at the DFT level; this would likely have the effect of reducing the energy differences between the ML and DFT estimates.
(Performing full relaxations at the DFT level, even with a highly scalable code like SPARC, would be extremely costly due to the scaling of DFT, the spin polarization, and the large number of degrees of freedom in the system.)
We performed a similar study on a smaller Co--Pt truncated octahedron \ce{Pt96Co105}, and the comparison between ML predicted putative minima against the L1$_0$ one is included in Figure S7 of SI.

\begin{figure}
\centering
\includegraphics[width=6.in]{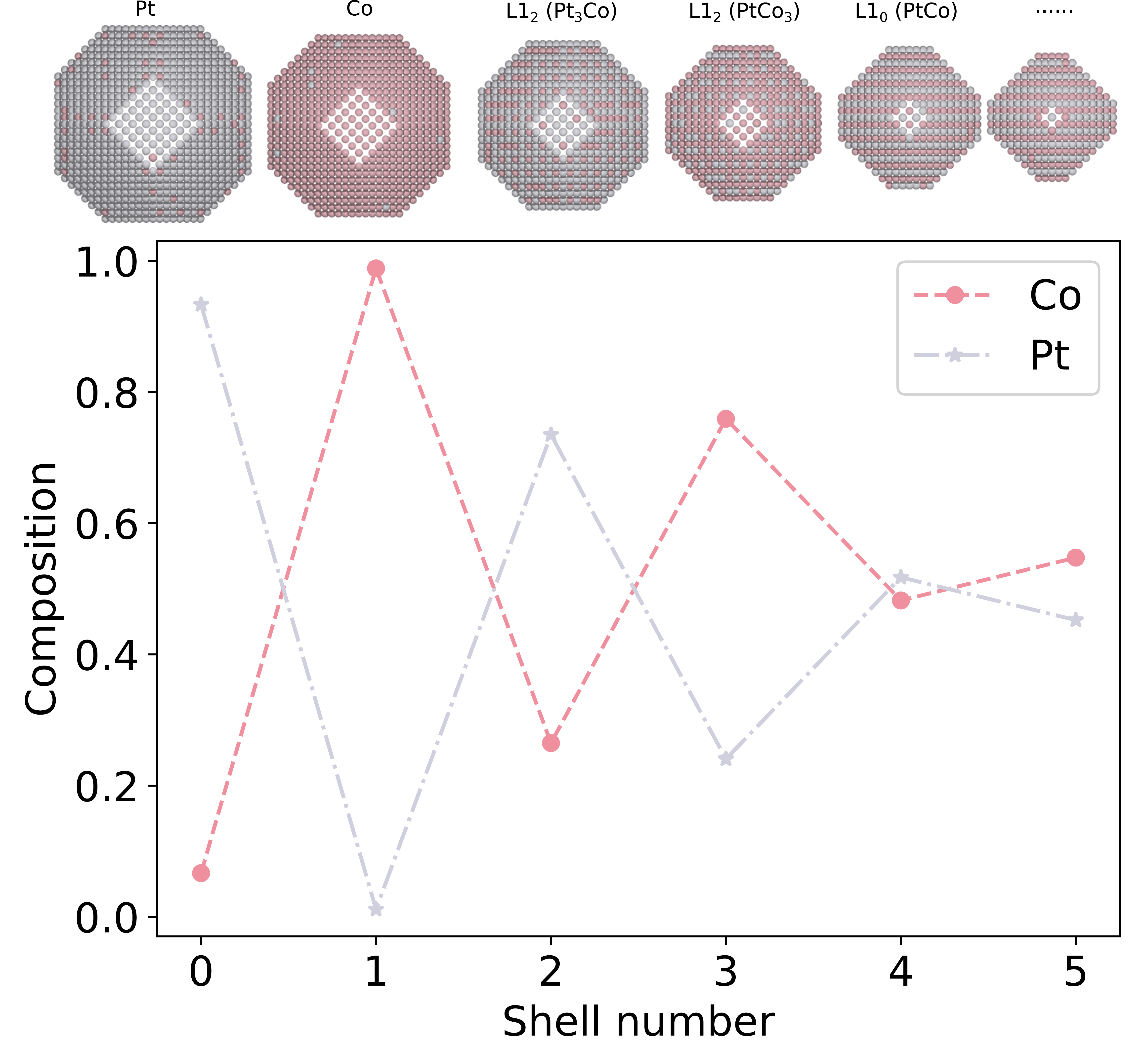}
\caption{Depth profile of compositions and configurations at each shell for a 6 nm \ce{Co3102Pt3164} optimized by Metropolis Monte Carlo simulations.}
\label{fig:mc-6266}
\end{figure}

We also performed Metropolis simulations at a temperature of 500 K on a much larger nanoparticle with 6266 atoms ($\sim$6 nm).
We started with a fully L1$_0$ ordered structure, and we would like to see where the thermodynamic fluctuations lead the structure.
We took out a structure after 330,000 steps, showing the configuration for each shell in Figure~\ref{fig:mc-6266}.
It is clear that that the first four outermost shells change to a distinct orderliness, whereas we find an almost unchanged L1$_0$-like structure going from the 5th shell to the center of the truncated octahedron.
Interestingly, the optimized structure for a truncated octahedron with nearly equal compositions of Pt and Co follows a pattern of atomic arrangement as:

\begin{equation*}\label{eq:new-toh-pattern}
	\ce{Pt \to Co \to $\mathrm{L}1_2$(Pt3Co) \to $\mathrm{L}1_2$(PtCo3) \to $\mathrm{L}1_0$(PtCo) \to \cdots \to $\mathrm{L}1_0$(PtCo)}
\end{equation*}

\noindent
So far, we can conclude that the optimal particle for a truncated octahedron with equal compositions displays concentric Pt and Co shells at the outermost two shells, then respective Pt-rich and Co-rich L1$_2$ ordered at the third and fourth shell, and fully L1$_0$ ordered close to the center.

\begin{figure}
\centering
\includegraphics[width=4.5in]{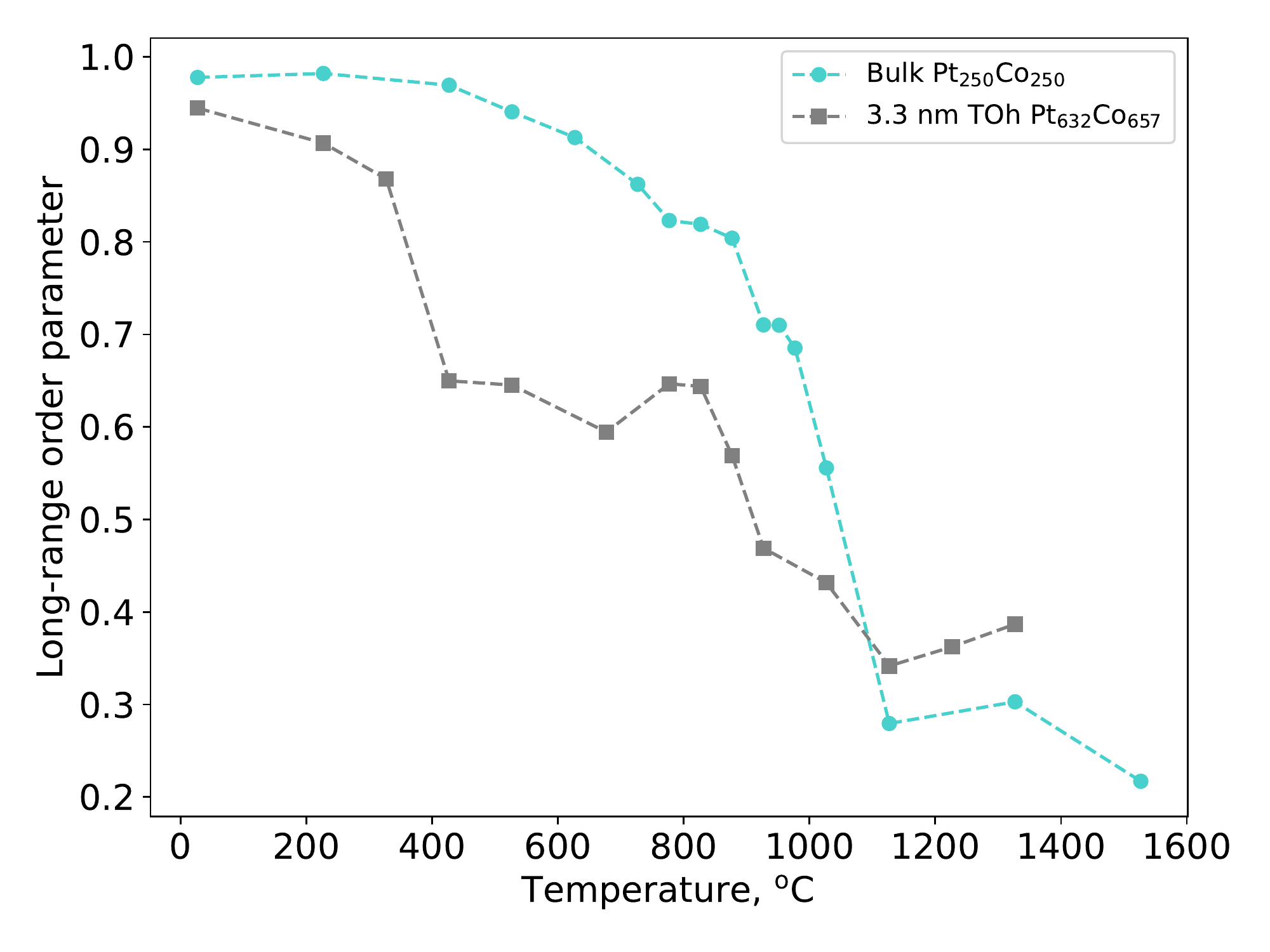}
\caption{Long-range order parameter for a 500-atom bulk cell and a 1289-atom 3.3 nm truncated octahedron calculated by Metropolis simulations at various temperatures. TOh is short for truncated octahedron.}
\label{fig:order--disorder}
\end{figure}

To find the temperature for order--disorder phase transition, we carried out a series of Metropolis simulations at various temperatures
We studied the order--disorder phase transition for two structures; one is a 500-atom bulk cell \ce{Pt250Co250} and the other is a 1289-atom truncated octahedron \ce{Pt632Co657}.
We employed a long-range order (LRO) parameter ($\Phi$) introduced by Cowley~\cite{Cowley1960} to describe the order--disorder transition, and it takes the form of:

\begin{equation}\label{eq:long-range-order}
\Phi = \max_{i \in \{x,y,z\}}(\{\Phi_i\})
\qquad
    \mbox{where}~\Phi_i = |p_{\mathrm{A}, i} - 1/2| + |p_{\mathrm{B}, i} - 1/2|
\end{equation}

\noindent
where $p_{\mathrm{A}, i}$ and $p_{\mathrm{B},i}$ are the occupation probabilities on each sublattice of the L1$_0$ phase evaluated in an ordering direction $i$.
As the stable structure of truncated octahedron found at 300 K is similar to that of the aforementioned 6266-atom structure, only exhibiting L1$_0$ ordering from the 5th shell to the center, we only consider those L1$_0$ ordered shells for the order--disorder transition in the 3.3-nm truncated octahedron nanoparticle \ce{Pt632Co657}.
Figure~\ref{fig:order--disorder} shows the order parameters calculated at various temperatures by Metropolis simulations on a bulk cell and a nanoparticle.
Although there isn't a sharp phase transition, we can see that the loss of order occurs in the vicinity of the experimentally observed transition temperature of $\sim$850$^\circ$C~\cite{Leroux1988, Bouar2003}, although our calculations appear to predict it to be slightly higher at a temperature of $\sim$1050$^\circ$C.
This deviation is similar to what has been found by well-validated empirical inter-atomic potentials. \cite{Alloyeau2009}
For the 1289-atom truncated octahedron, there exists a much smoother transition region where the transition temperature is found to be around 900$^\circ$C, which is 150$^\circ$C lower than that of a bulk.
The size effect agrees well with experimental observations and simulations in the work of Alloyeau \etal \cite{Alloyeau2009}, in which order--disorder phase transition temperature is lowered by at least 175 $^\circ$C.
This phenomenon can be understood by the surface induced disordering due to the reduced coordination hence an overall lowered order--disorder transition temperature~\cite{Yang2006}.

\section{Conclusions}

Based on an NFT approach, robust neural network models have been developed for \CoPt\ nanoparticles of up to several thousand atoms in size, using training sets containing images with no more than 168 atoms/image.
This work also demonstrates that the NFT approach is applicable to multi-element magnetic nanoparticles.
The resulting models can readily be improved by addressing uncertain local chemical environments when necessary.
By pairing these models with genetic algorithms and Metropolis Monte Carlo simulations, we have presented a thorough study of the stable structures of Co--Pt nanoparticles.
We summarize the key findings below, which not only refine existing understandings of the thermodynamic stability of Co--Pt nanoparticles, but also offer guidelines for the synthesis of nanoparticle catalysts in experiments.
The experimental guidelines include, but are not limited to, using temperature to control the orderliness of the nanoparticle and tuning surface energy with capping agent targeting desired nanoparticle shapes.

\begin{enumerate}

    \item \CoPt\ nanoparticles exhibit a strong tendency to form alternating layers near the surface, with a platinum-rich skin and a cobalt-rich underlayer.
This was seen in many systems throughout this study, and the concentric nature continued through the 4th shell in the case of a 6-nm (6266-atom) structure. 

  \item \CoPt\ fcc(100) surfaces also exhibit a strong tendency to form an L1$_0$ ordered structure featuring alternating Pt and Co layers.
  \CoPt\ fcc(111) surfaces show more flexibility of the atomic arrangement while the major feature is also the alternating layers.

  \item
The truncated octahedron is the most stable shape for Pt nanoparticles of moderate to large sizes (200--7000 atoms), due to its low surface and volume energies
This explains its frequent appearance in experiments.
The stability of icosahedron and cuboctahedron particles is always less, but these two shapes exhibit a crossover in stability at a size of $\sim$500 atoms.

\item
The truncated octahedron is the most stable shape for large \CoPt\ nanoparticles, while the icosahedron is more stable for smaller nanoparticles.
An composition-dependent empirical model was introduced to study the crossover among structural motifs in Co--Pt nanoparticles.
The addition of Co improves the stability of icosahedron, leading to a crossover between icosahedron and truncated octahedron at the size of $\sim$333 atoms for a given Co composition of 35\%.
It can be rationalized by the significant stress release on the distorted fcc(111) surfaces of icosahedron when a smaller element is introduced in the core.
The crossover moves to a larger size when more Co atoms are added.

  \item Metropolis simulations reveal that the most stable atomic arrangement of a Co--Pt truncated octahedron with nearly equal Co and Pt compositions is not fully L1$_0$ ordered, as often found by well-parameterized empirical potentials.
  Instead, it displays a more complex pattern going from the outermost shell to the center of truncated octahedron, which is confirmed by large-scale DFT calculations on SPARC. The outermost shell is Pt segregated, followed by a Pt depleted second shell. The third and fourth shells are L1$_2$-like structures rich in Pt and Co, respectively.
  Deeper shells all exhibit L1$_0$-like atomic arrangement.

  \item The order--disorder phase transition for a bulk and nanoparticle has been studied  based on a long-range order parameter.
  Nanoparticles show a lower transition temperature and a much smoother transition compared to a bulk Co--Pt alloy.

  \item The energy convex hull for a 147-atom Co--Pt icosahedron constructed by neural network models is quantitatively accurate compared to brute-force \abin~ calculations, and a new low-energy atomic arrangement for \ce{Pt80Co67} is identified.

\end{enumerate}

\subsubsection*{SUPPORTING INFORMATION}
Representative cubocahedron and icosahedron structures used to generate the training data (atomic chunks) and the corresponding average atomic uncertainties for the relaxed structures of those full-size nanoparticles (S-1), statistics for the atomic chunks (S-2), validation of force consistency between GPAW and SPARC (S-3), energies and configurations for a 147-atom icosahedron and cuboctahedron (S-4), energies and configurations for two \ce{Pt80Co67} icosahedra (S-5), fitted size dependency of pure Pt nanoparticle energies including edge terms (S-6), and shell-by-shell compositions and configurations of a \ce{Pt96Co105} truncated octahedron (S-7) are included in the supporting information.

\subsubsection*{SUPPORTING DATASET}
The supporting dataset is saved in a private repository and will be made available to public after this work is accepted for publication. Scripts to reproduce some simulations will also be provided along with the supporting dataset.

\subsubsection*{AUTHOR INFORMATION}

\paragraph{Corresponding author.}
*Email: andrew\_peterson@brown.edu. Telephone +1 401-863-2153.

\paragraph{Acknowledgments.}
The authors thank Phanish Suryanarayana and Qimen Xu for their support in running the SPARC code and providing feedback on convergence issues.
This work was supported by the U.S.\ Department of Energy's Basic Energy Science, Computational Chemical Sciences Program Office, under Award No. DE-SC0019441.
Cheng Zeng acknowledges support from the Presidential Fellowship at Brown University.
This work was undertaken using the computational resources at the Brown University Center for Computation and Visualization (CCV).


\bibliography{acs-catal}
\bibliographystyle{jpclp}

\end{document}


\definecolor{keywords}{RGB}{255,0,90}
\definecolor{comments}{RGB}{0,0,113}
\definecolor{red}{RGB}{160,0,0}
\definecolor{green}{RGB}{0,150,0}

\lstset{language=Python,
        basicstyle=\ttfamily\small,
        keywordstyle=\color{keywords},
        commentstyle=\color{comments},
        stringstyle=\color{red},
        showstringspaces=false,
        identifierstyle=\color{green},
        procnamekeys={def,class}}
\maketitle
\thispagestyle{fancy}



\newcommand{\callout}[1]{\textbf{#1}}

This Supporting Information (SI) includes the representative cubocahedron and icosahedron structures used to generate the training data (atomic chunks) and the corresponding average atomic uncertainties for the relaxed structures of those full-size nanoparticles, statistics for the atomic chunks, validation of force consistency between GPAW and SPARC, energies and configurations for a 147-atom icosahedron and cuboctahedron, energies and configurations for two \ce{Pt80Co67} icosahedra (one is of corner occupancy and the other is of terrace center occupancy on the surface), fitted size dependency of pure Pt nanoparticle energies including the edge terms, and shell-by-shell compositions and configurations of a \ce{Pt96Co105} regular truncated octahedron.

\section{Nanoparticles used to generate training data}

\begin{figure}
\centering
\includegraphics[width=5.5in]{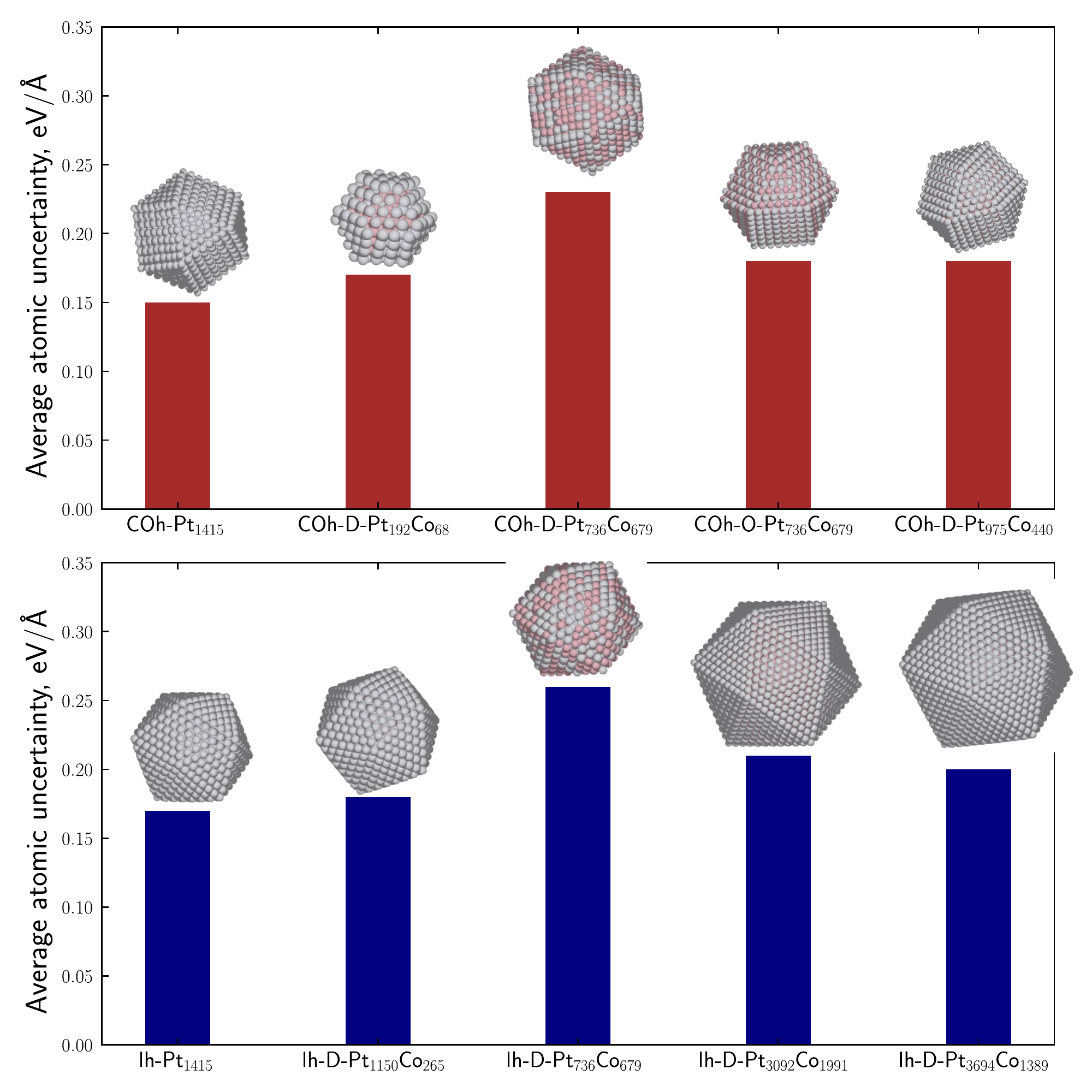}
\caption{Nanoparticles used to generate atomic chunks and the corresponding average atomic uncertainty of forces of their relaxed structures. COh and Ih represent cuboctahedron and icosahedron, respectively. 'D' and 'O' indicate respective disordered and ordered structures.}
\label{fig:structure_motifs}
\end{figure}

Figure.~\ref{fig:structure_motifs} shows average atomic uncertainties of forces for the relaxed structures of representative nanoparticles.
The corresponding configurations are shown at the top of each bar.

\section{Statistics of generated atomic chunks}
\begin{figure}
\centering
\includegraphics[width=5.5in]{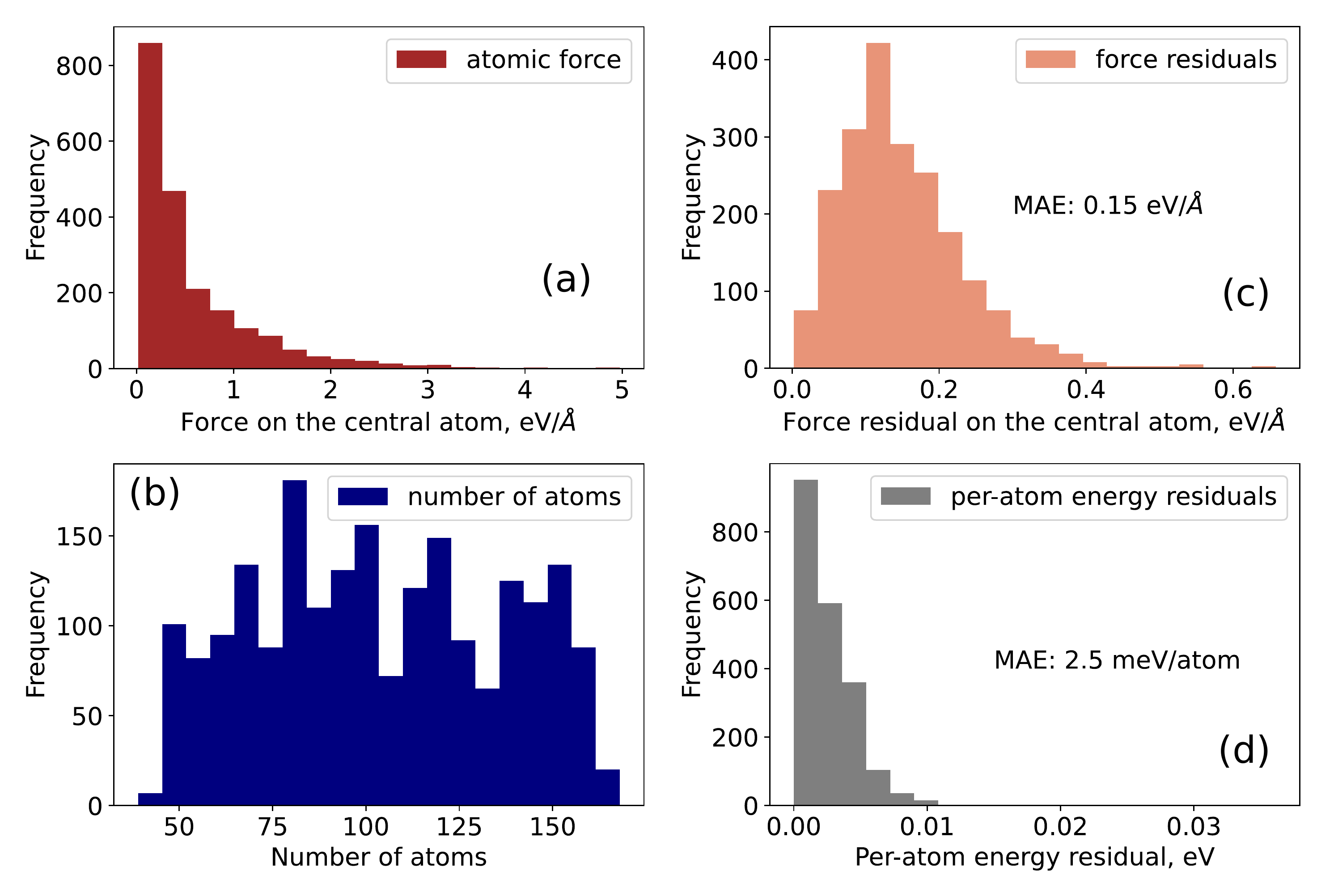}
\caption{Statistics of generated atomic chunks. (a) Forces on the central atom, (b) Number of atoms in the atomic chunk, (c) Force prediction residuals using the force ensemble model, (d) Per-atom energy prediction residuals using the energy ensemble model.}
\label{fig:chunk_stats}
\end{figure}

Figure.~\ref{fig:chunk_stats} shows statistics of the atomic chunks, including DFT-calculated atomic forces on the central atoms (a), number of atoms (b), force prediction residuals (c), and per-atom energy prediction residuals (d).
The average force magnitude is calculated to be 0.59 eV/\AA.
The average number of atoms is around 104 atoms.
The mean absolute deviation (MAE) of atomic forces between DFT forces and ML-predicted forces is 0.15 eV/\AA.
The MAE of per-atom energy between DFT energies and ML-predicted energies is 2.5 meV/atom.

\section{Validation of force consistency between GPAW and SPARC}
\begin{figure}
\centering
\includegraphics[width=4.5in]{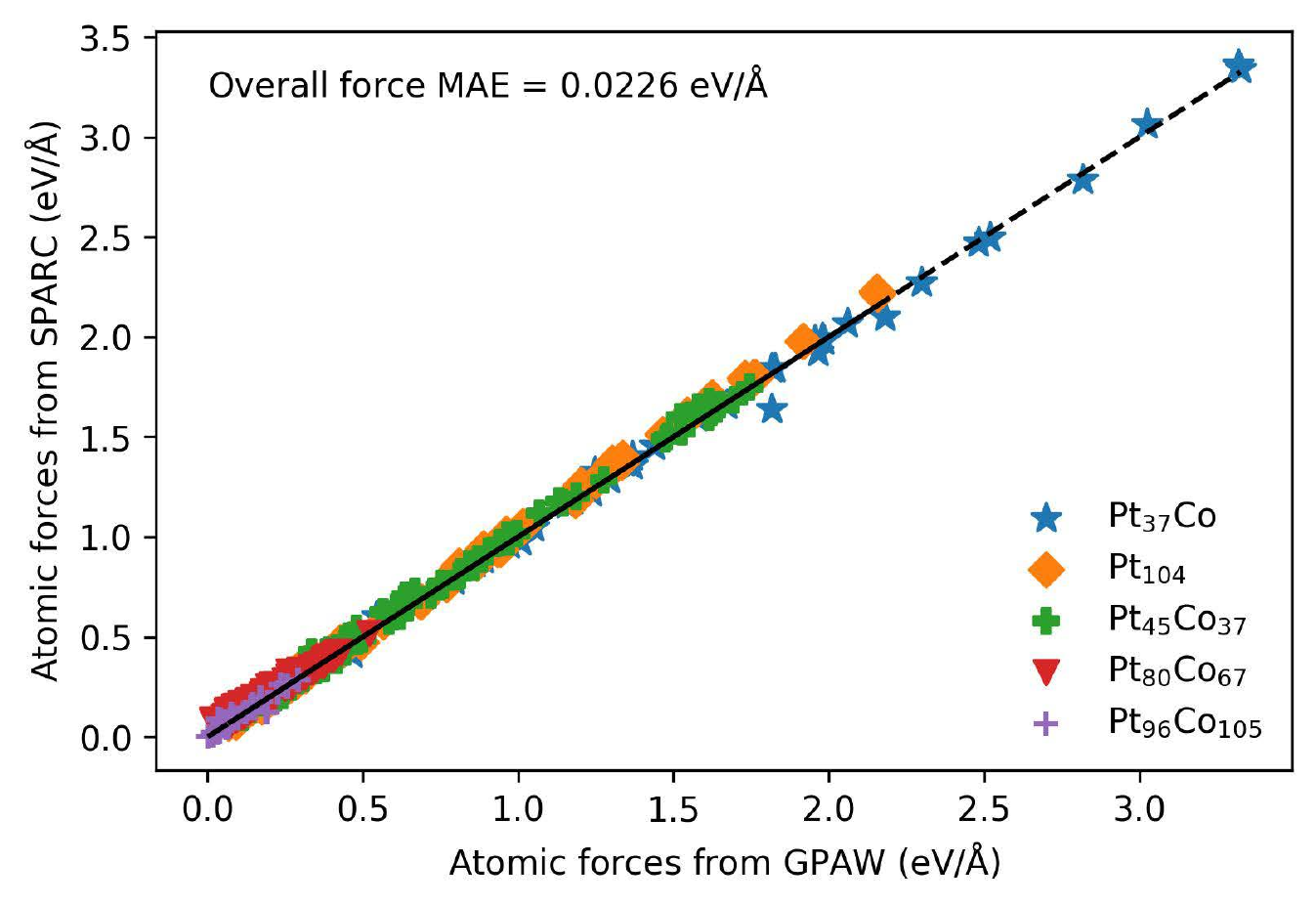}
\caption{Comparison of forces from GPAW and SPARC on five test clusters, including \ce{Pt37Co}, \ce{Pt104}, \ce{Pt45Co37}, \ce{Pt80Co67} and \ce{Pt96Co105}.}
\label{fig:force_validation}
\end{figure}

This work is mainly concerned with atomic forces and relative energy differences between different nanoparticles/clusters.
We used two DFT codes with different computational settings, namely GPAW and SPARC.
To make sure that both codes lead to consistent force predictions,
we compare the atomic forces from GPAW and SPARC on various systems, as shown in the Figure.~\ref{fig:force_validation}.
It is clear that small MAEs are observed across all systems, leading to an overall MAE of around 0.023 eV/\AA.
It confirms the consistency of force predictions between GPAW and SPARC.

\section{Energies and configurations for a 147-atom icosahedron and cubocahedron}
\begin{figure}
\centering
\includegraphics[width=4.5in]{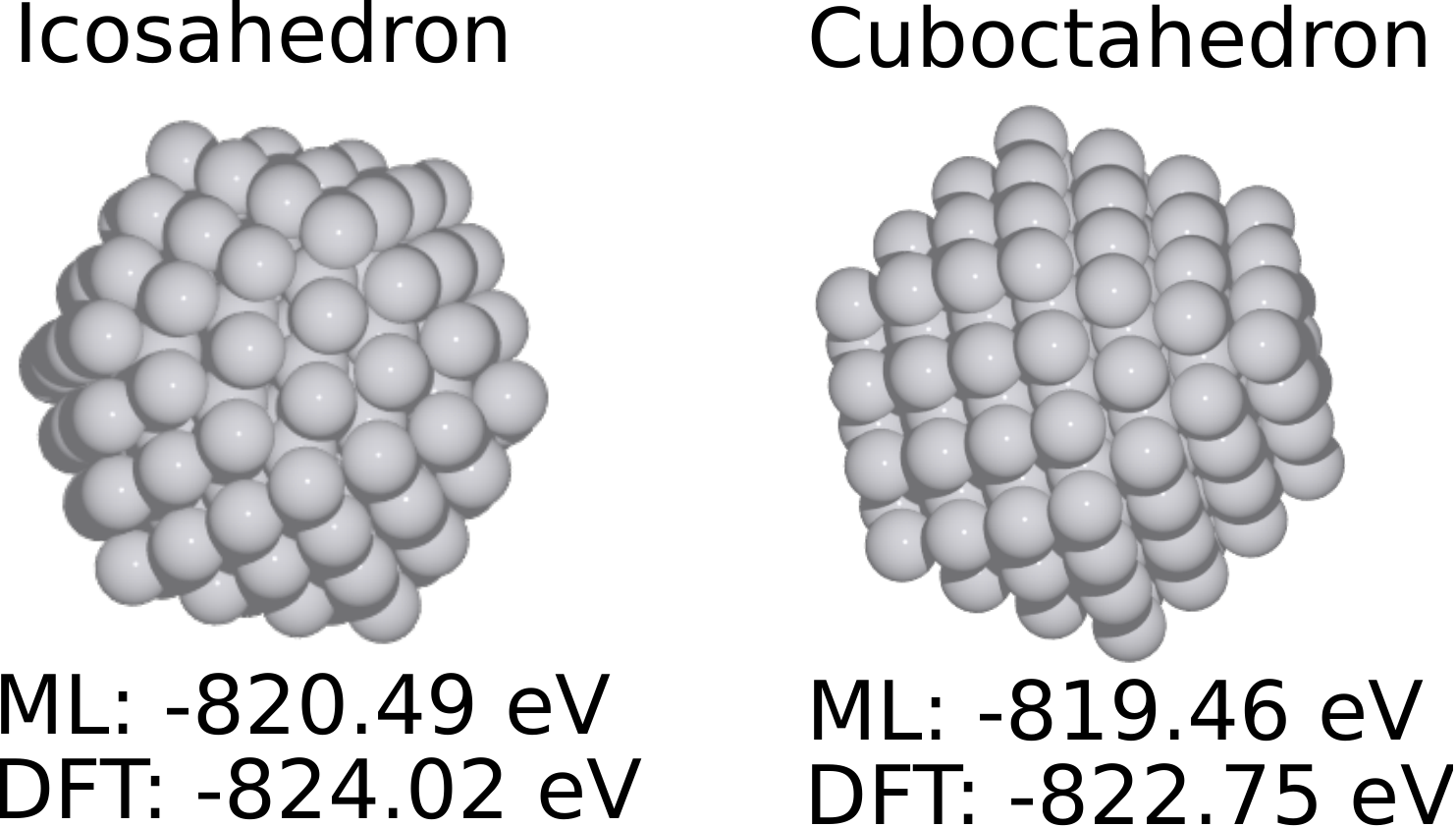}
\caption{Configurations and energies of a 147-atom icosahedron and cubocahedron.}
\label{fig:Ih_vs_COh}
\end{figure}

Figure.~\ref{fig:Ih_vs_COh} displays the ML-predicted and DFT-calculated engergies for a 147-atom icosahedron and cubocahedron relaxed by ML models.
Both ML models and DFT calculations show that the icosahedron structure is more stable than the cubocahedron structure by around 1 eV.

\section{Energies and configurations for two types of \ce{Pt80Co67} icosahedra}
\begin{figure}
\centering
\includegraphics[width=4.5in]{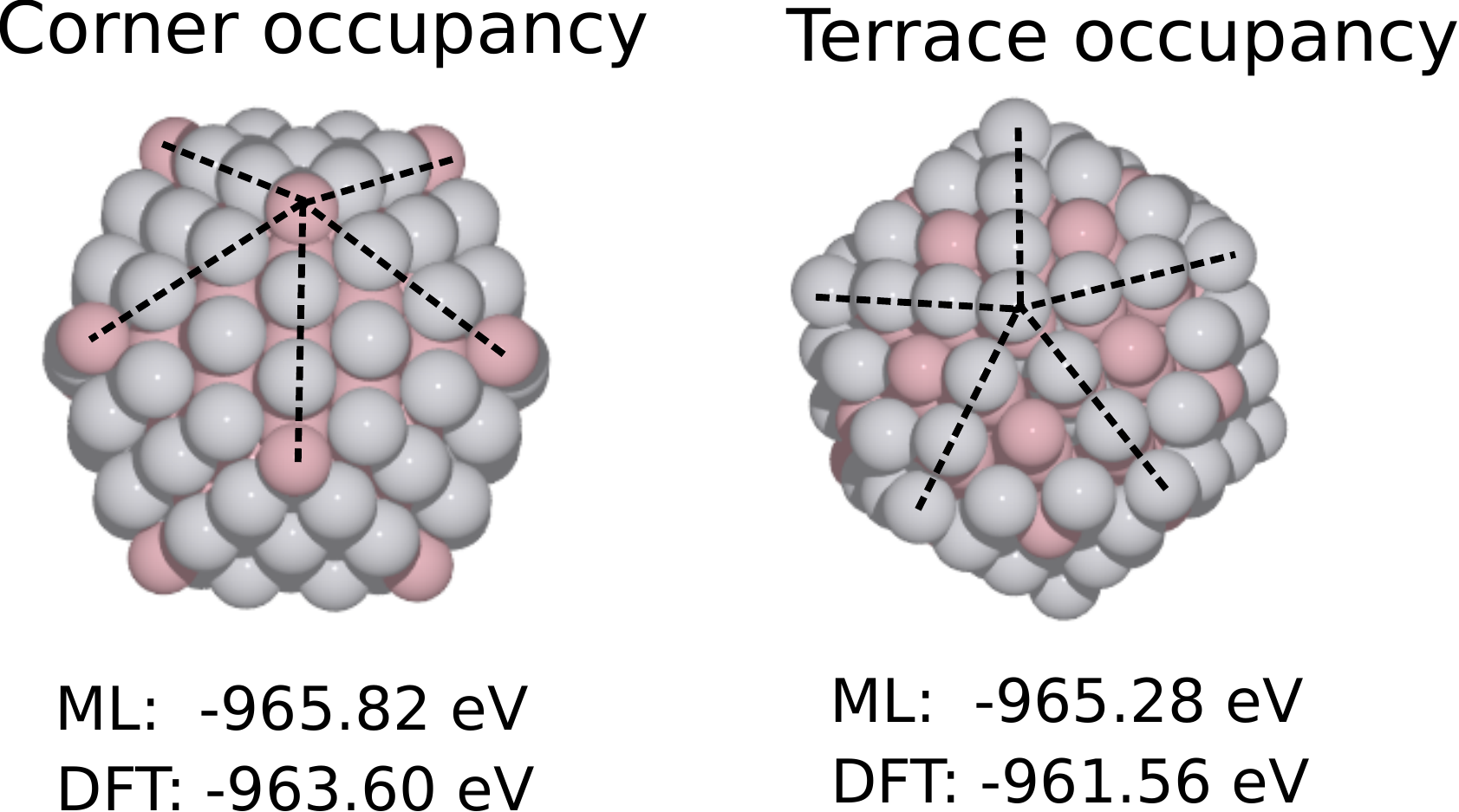}
\caption{Configurations of two types of \ce{Pt80Co67}, with Co atoms on the surface occupying the corner (Left) and occupying the terrace center (Right).}
\label{fig:terrace_vs_corner}
\end{figure}

Figure.~\ref{fig:terrace_vs_corner} shows two low-energy configurations of \ce{Pt80Co67}.
The one with corner occupancy is identified by brute-force DFT calculations~\cite{Noh2013}.
The one with terrace center occupancy is found by the GA study using ML models.
Both ML models and DFT calculations suggest that corner occupancy is more energetically favorable than terrace center occupancy by $\sim$4 meV/atom and $\sim$14 meV/atom, respectively.
However, the small DFT energy difference between two structures implies that the low-energy configuration of terrace center occupancy is missing in the brute-force approach~\cite{Noh2013}.

\section{Fitted size dependency of pure Pt nanoparticle energies including the edge terms}

\begin{figure}
\centering
\includegraphics[width=5.0in]{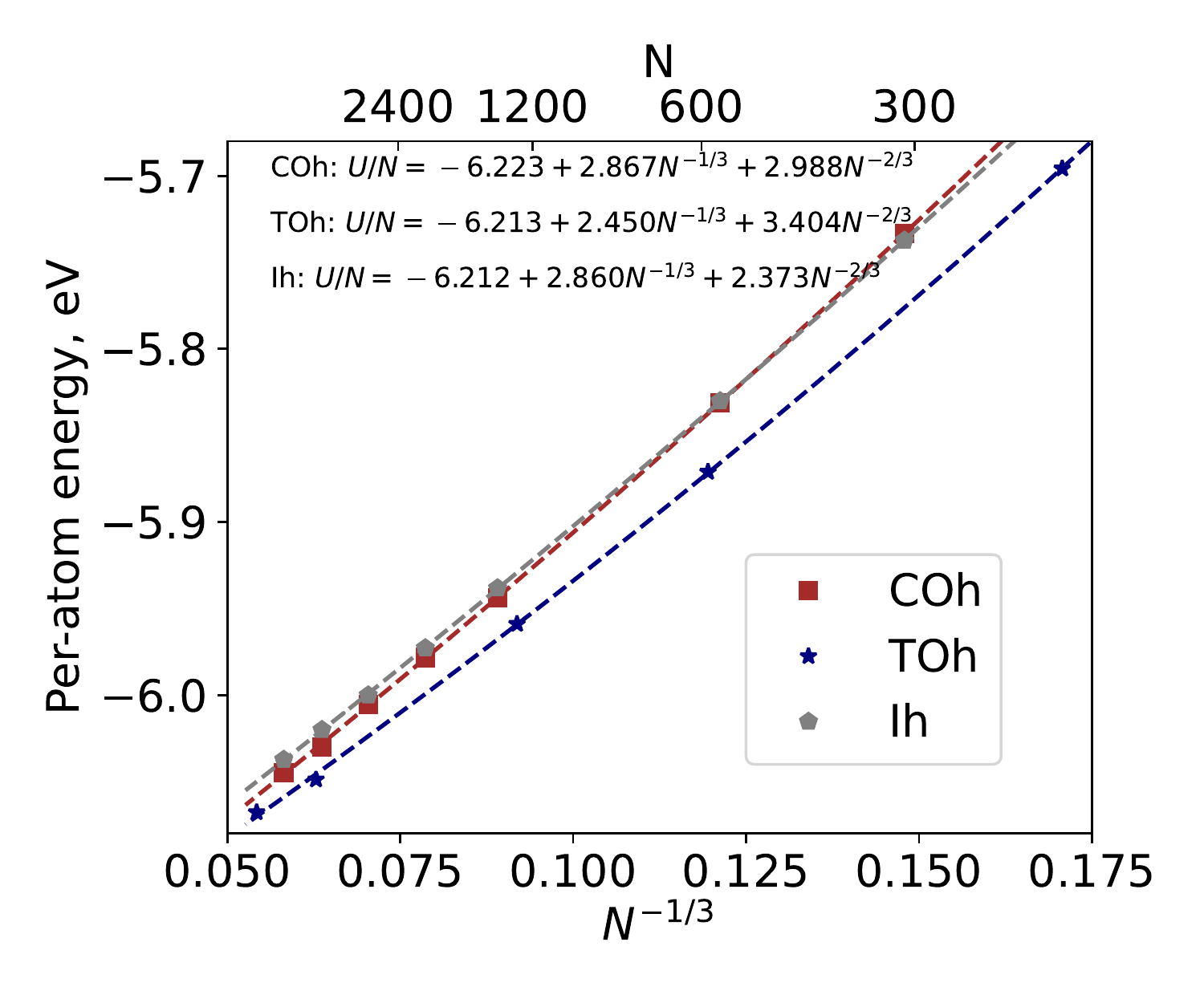}
\caption{Energies of relaxed structure motifs of Pt nanoparticles, plotted as per-atom energy ($U/N$) versus $N^{-1/3}$.
COh, TOh and Ih represent cuboctahedron, truncated octahedron and icosahedron, respectively.
The fitted results were based on Eq. 2 where edge terms were considered.}
\label{fig:pt-crossover-with-edge}
\end{figure}

Figure~\ref{fig:pt-crossover-with-edge} shows the raw ML predicted energetics and the fitted results using the Eq. 2. At a first glimpse, the coefficient for edge contributions are comparable to that for surface contributions, but if we take into account the value of $N^{-1/3}$, the edge term becomes much less significant.

\section{Shell-by-shell compositions and configurations of two \ce{Pt96Co105} truncated octahedron}
For a \ce{Pt96Co105} truncated octahedron, MC simulations at 300 K were performed to find the putative global minima.
A structure after ~9200 MC steps was regarded as the MC minima.
It was compared to the fully L1$_0$ ordered structure.
Both structures were relaxed by the force model and the energetics were obtained by the energy model, and a comparison for the energy difference was made with the DFT result.
Both ML models and DFT calculations find the MC structure to be more stable than the L1$_0$ structure, by energy differences of 17.4 eV (0.086 eV/atom) and 22.3 eV (0.111 eV/atom), respectively.
The DFT calculated maximum atomic forces for the MC and L1$_0$ structures are 0.41 and 0.29 eV/\AA, respectively, both of which are within the maximum atomic uncertainties by ML models, \textit{i.e.} 0.5 and 0.45 eV/\AA.
It suggests that the most stable atomic arrangement of a Co--Pt truncated octahedron with nearly equal compositions is not fully L1$_0$ ordered.

\begin{figure}
\centering
\includegraphics[width=6in]{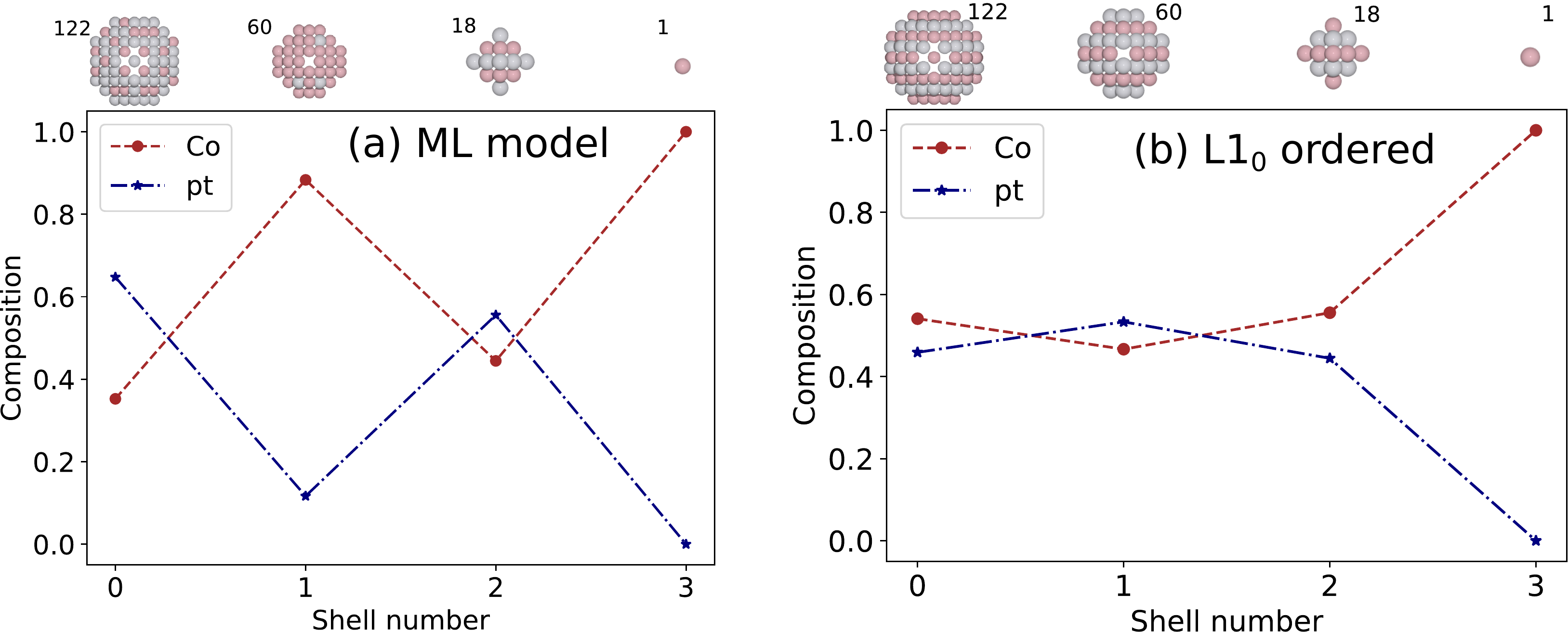}
\caption{Composition depth profile of a truncated octahedron Pt$_{96}$Co$_{105}$:  the putative global minima found by ML models (a), and the fully L1$_0$ ordered Co--Pt nanoparticle alloy (b). Atomic arrangement at each shell and the total number of atoms are provided at the top.}
\label{fig:co105pt96_mc_vs_l10}
\end{figure}

\singlespace
\bibliography{acs-catal}
\bibliographystyle{jpclp}